\documentclass[preprints,article,accept,moreauthors,pdftex]{mdpi} 

\setitemize{parsep=6pt,itemsep=0pt,leftmargin=*,labelsep=5.5mm,align=parleft}
\setenumerate{parsep=6pt,itemsep=0pt,leftmargin=*,labelsep=5.5mm,align=parleft}
\usepackage{amssymb}
\usepackage{soul}

\firstpage{1} 
\pubvolume{xx}
\issuenum{1}
\articlenumber{5}
\pubyear{2020}
\copyrightyear{2020}
\history{{Received: 16 June 2020; Accepted: 17 July 2020; Published: date}}

\newcommand{\be}{\begin{equation}}
\newcommand{\ee}{\end{equation}}
\newcommand{\beq}{\begin{eqnarray}}
\newcommand{\eeq}{\end{eqnarray}}
\newcommand\subsun[1]{{$_{\normalsize\odot}$}}

\def\prd{Phys.~Rev.~D}%
\def\prl{Phys.~Rev.~Lett.}%
\def\sovast{Soviet~Ast.}%
\def\lsim{\;\raise0.3ex\hbox{$<$\kern-0.75em\raise-1.1ex\hbox{$\sim$}}\;}
\def\gsim{\;\raise0.3ex\hbox{$>$\kern-0.75em\raise-1.1ex\hbox{$\sim$}}\;}

\def\prd{Phys.~Rev.~D}%
\def\prl{Phys.~Rev.~Lett.}%
\def\sovast{Soviet~Ast.}%
\Title{The Origin of Matter at the Base of Relativistic Jets in Active Galactic Nuclei}

\Author{Gustavo E. Romero $^{1,2,}$*$^{,\dagger}$\orcidA{} and Eduardo Guti\'errez  $^{1}$\orcidB{}}   

\AuthorNames{Gustavo E. Romero \& Eduardo Gutiérrez}

\address{%
$^{1}$ \quad Instituto Argentino de Radioastronom\'{\i}a (CONICET; CICPBA), C.C. No. 5, Villa Elisa 1894, Argentina; egutierrezposse@gmail.com\\ 
$^{2}$ \quad Facultad de Ciencias Astron\'omicas y Geof\'{\i}sicas, Universidad Nacional de La Plata, Paseo del Bosque s/n, La Plata 1900, Argentina}

\corres{Correspondence: romero@iar-conicet.gov.ar or romero@fcaglp.unlp.edu.ar}  

\firstnote{Member of CONICET.} 

\abstract{The generation of relativistic jets in active sources such as blazars is a complex problem with many aspects, most of them still not fully understood. Relativistic jets are likely produced by the accretion of matter and magnetic fields onto spinning black holes. Ergospheric dragging effects launch a Poynting-dominated outflow in the polar directions of these systems. Observations with very high resolution of the jet in the nearby radio galaxy M87 and evidence of extremely fast variability in the non-thermal radiation of several other objects indicate that charged particles produce synchrotron emission and gamma rays very close to the base of the jet. How these particles are injected into the magnetically shielded outflow is a mystery. Here we explore the effects of various processes in the hot accretion inflow close to the black hole that might result in the copious production of neutral particles which, through annihilation and decay in the jet's funnel, might load the outflow with mass and charged particles on scales of a few Schwarzschild radii. }


\begin{document}

\setcounter{section}{0} 

\section{Introduction}
\label{sec:Introduction}

Astrophysical jets are collimated outflows of fields and particles. They are observed on a variety of scales and situations, from protostars to supermassive black holes (SMBHs) in Active Galactic Nuclei (AGNs).  The basic ingredients for the production of such jets seem to be accretion onto a gravitating object and magnetic fields \cite{Spruit2010}.  The most extreme examples of jets occur in AGNs where a spinning supermassive black hole is fed through a hot accretion disk. The power of the jet is directly related to the accretion power $\dot{M}c^2$, where the $\dot{M}$ is the mass accretion rate and $c$ is the speed of light in vacuum, by $P_{\rm jet}=q \dot{M} c^2$ \cite{FalckeBiermann1995,RomeroVila2014}. In galactic microquasars such as Cyg X-1, $q\sim0.1$ \cite{Pepeetal2015}. Some extragalactic sources, on the other hand, can be far more efficient. In particular,  Flat Spectrum Radio Quasars (FSRQs) and other types of blazars can reach much higher values of $q$; in many cases even with $q>1$. This~means an efficiency of more than 100\%, which requires direct extraction of energy from the black hole. Such~a~situation becomes possible when the magnetic pressure close to the black hole exceeds the pressure of the infalling gas and the disk becomes magnetically arrested (MAD, for Magnetically Arrested Disk,  \cite{Bisnovatyi-Kogan1974,Narayanetal2003,Tchekhovskoyetal2011,McKinneyetal2012}). The order of magnitude of the different physical parameters that characterize the situation can be estimated as follows.

The magnetically dominated region is determined by the condition that the attractive gravitational force is balanced by the force due to the magnetic pressure \cite{Tchekhovskoy2015}:
\beq
\frac{B^2}{8\pi} 4 \pi r^2=\frac{G M_{\rm BH} \rho 4 \pi r^3/3}{r^2}, \label{MAD}
\eeq
where $B$ is the magnetic field, $r$ is the radius measured from the black hole center, $M_{\rm BH}$ the black hole mass, and $\rho$ the density of the accreting mass, which relates to the accretion rate and the height $h $ of the disk by the continuity equation:
\beq
\dot{M}=4 \pi r^2 \rho v_r \left( \frac{h}{r} \right).\label{Cont}
\eeq
Here, $v_r$ is the radial\footnote{Radial distances are measured perpendicularly to the jet axis throughout the paper. When we refer to the accretion disk the radial distances are along the equatorial plane of the disk.} velocity of the gas toward the black hole. 

From Equations (\ref{MAD}) and (\ref{Cont}) we obtain the magnetic field in the MAD region \cite{Tchekhovskoy2015}:
\beq
B_{\rm MAD}\sim 0.4\times 10^4 \left( \frac{L}{0.1 L_{\rm Edd}}\right)^{1/2}\; \left(\frac{M}{10^9 M_{\odot}} \right)^{-1/2} \left(\frac{h}{r}\right)^{-1/2}\;\; \textrm{G},
\eeq
where $L_{\rm Edd} \approx 1.2\times 10^{38} (M_{\rm BH}/ M_{\odot})$   erg s$^{-1}$ is the Eddington luminosity and we have assumed that the infall velocity is $v_r \sim c$. 

The event horizon of a Kerr black hole with  dimensionless spin parameter $a$ is located at \cite{RomeroVila2014}: 
\beq
r_{\rm H}= r_{\rm g} \left( 1 + \sqrt{1-a^2}\right), \label{rH}
\eeq
with $ r_{\rm g}=GM_{\rm BH}/c^2$ the gravitational radius. The angular velocity of the horizon measured by an observer at infinity is
\beq
\Omega_{\rm H}=\frac{ac}{2r_{\rm H}}. \label{OmegaH}
\eeq
The magnetic field lines frozen in the plasma are dragged with an angular velocity that goes from $\Omega_{\rm H}$ close to the horizon to zero far away. On average, the dragging velocity is $\Omega_B\sim 0.5 \Omega_{\rm H}$ \cite{BlandfordZnajek1977,Tchekhovskoyetal2010}.  

The accreted field lines in the MAD region are essentially poloidal\footnote{The magnetic field can be decomposed as $\textbf{B} = \textbf{B}_{\rm p} + B_\phi \hat{\phi}$, where  $\textbf{B}_{\rm p} \equiv B_r \hat{r}+B_z \hat{z}$ is the \emph{poloidal} component of the field and $B_\phi$ is the \emph{toroidal} component.}.  The dragging of this field results in the development of a toroidal field and a Poynting flux directed along the rotation axis away from the black hole. The power of this flux is

\beq
P\sim \frac{c}{4\pi} (\textbf{E}\times \textbf{B})_r \times 4 \pi r_{\rm lc}^2  \label{P1},
\eeq
where $r_{\rm lc}$ is the radius of the light cylinder, which gives the approximate size of the MAD region: $r_{\rm lc}=c/\Omega_B$.
We can introduce now a normalized $\Omega_{\rm H}$ \cite{Tchekhovskoy2015}:

\beq
\omega_{\rm H}\equiv \frac{2r_{\rm g}\Omega_{\rm H}}{c}=\frac{a}{1+\sqrt{1-a^2}}.
\eeq

Since all components of the electromagnetic field become of the same order in this region, Equation~(\ref{P1}) can be rewritten as (see Refs. \cite{Tchekhovskoy2015,Tchekhovskoyetal2010} for details):

\beq
P\sim \frac{c}{64 \pi^2 r_{\rm g}^2} \Phi_{\rm BH}^2 \omega_{\rm H}^2 \;  \label{P},
\eeq
where $\Phi_{\rm BH}$ is the magnetic flux through the black hole. 

Making the 6th order expansion we get \cite{Tchekhovskoyetal2010}:
\beq
P\sim \frac{G^2}{8 c^3} B^2 M_{\rm BH}^2 \; f(a),  \label{P2}
\eeq
where we used Equations (\ref{rH}) and (\ref{OmegaH}), and the definitions of $r_{\rm g}$ and $r_{\rm lc}$. The function $f(a)$ tends to $a^2$ for low $a$ (see Figures 3 and 4 in Ref. \cite{Tchekhovskoy2015}), yielding the well-known Blandford-Znajek formula: 

\beq
P_{\rm BZ}\propto B^2 M_{\rm BH}^2 a^2  \label{PBZ1},
\eeq
or, in convenient units,
\begin{equation}
 P_{\rm BZ}\approx 10^{46}\,\left(\frac{B}{10^4 \rm G}\right)^2 \left(\frac{M}{10^9 M_\odot}\right)^2a^2\,\,\rm{erg\,s}^{-1}.
\label{PBZ2}
\end{equation}
The latter expression is accurate for $a\leq 0.5$. Beyond this limit, the formula under-predicts the true jet power by a factor of $\approx 3$ and the 6th order expansion of $f(a)$ is necessary. For details see Ref. \cite{Tchekhovskoyetal2010}. In~any case, Equation (\ref{PBZ2}) shows that a rapidly rotating black hole in a MAD regime can produce a Poynting flux dominated jet with an efficiency $\eta_{\rm BZ}=P_{\rm BZ}/\dot{M}c^2 \gg 1$. 

 An important problem to deal with is that of how this electromagnetic outflow can be loaded with mass immediately after its generation in the ergosphere. Several authors have tackled this question and proposed various mechanisms for particle injection: pair creation via the collisions of MeV photons \cite{Moscibrodzkaetal2011,LevinsonRieger2011} and/or more energetic gamma-ray  photons \cite{KimuraToma2020} produced in the accretion flow, or via electromagnetic cascades in the polar region \cite{HirotaniOkamoto1998,BroderickTchekhovskoy2015}, and hadronic injection via neutron decay \cite{Kimuraetal2014,Toma2012,Vilaetal2014}. In this paper, we discuss quantitatively the relative contributions from each of these mechanisms. We also explore in much detail the role that non-thermal processes within the accretion flow onto the central engine might play on the loading of jets.

In Section \ref{sec2} we state the nature of the problem dealt with. Section \ref{sec3} is devoted to a characterization of the black hole environment, including the structure of the accretion flow and the associated radiation. Section \ref{pairs} discusses the different mechanism for pair injection at the base of the jet. Section \ref{baryons} deals with the injection of baryons. The total injection is discussed in Section \ref{sec:total}. Finally, Sections \ref{sec7} and \ref{sec8} present some additional discussions and our conclusions.

\section{The Problem}\label{sec2}

The basic features of the magnetic model for the generation of relativistic jets outlined in the Introduction are  the following \cite{Koideetal2000,Koideetal2002,Meieretal2001,Tchekhovskoy2015}:

\begin{itemize}
 \item Relativistic jets are produced by rapidly rotating BHs fed by magnetized accretion disks.
\item The ultimate power source is the rotational energy of the black hole.
\item The energy is extracted via magnetic torque as Poynting flux.
\item Jet collimation is due to the external medium outside the MAD zone\footnote{Magnetic self-collimation due to the toroidal component of the field is not effective in relativistic plasma flows due to kink instabilities \cite{Eichler1993}, so confinement by the pressure and inertia of an external medium seems to be quite essential \cite{GlobusLevinson2016}.}.
\item Jet acceleration is via conversion of the electromagnetic energy into bulk kinetic energy.
\end{itemize}

A torsional Alfv\'en wave is generated by the rotational dragging of the poloidal field lines near the black hole.  The field develops a toroidal component perturbing the field. The wave transports magnetic energy outward, causing the total energy of the plasma near the hole to decrease to negative values. When this negative energy plasma enters the horizon, the rotational energy of the black hole decreases. Through this process, the energy of the spinning black hole is extracted magnetically. The~result of all this is that while plasma is carried into the hole only (not ejected), electromagnetic power is evacuated along the rotation axis. Notice that the magnetic field is tied to the infalling plasma, not to the horizon, as in the simplified split monopole picture \cite{BlandfordZnajek1977}. The back-reaction of the magnetic field accelerates the ergospheric plasma to relativistic speeds counter to the hole's rotation endowing the plasma with negative energy. It is the accretion of this negative energy plasma which spins down the~hole. 

Despite the fact that the evacuation funnel around the rotation axis of the black hole is expected to be free of plasma, very-high-resolution images of the central source of the nearby AGN M87 obtained with the Event Horizon Telescope (EHT) \cite{ETH-1-2019} at 230 GH and Very Long Baseline Array data at 43 and 86 GHz \cite{Hada2013} reveal the existence of emission of radiation associated with the jet at distances down to $\sim$10 gravitational radii from the SMBH \cite{Hada2017}. This radiation is produced by relativistic electrons or positrons injected and accelerated somehow in the region. The jet also accelerates from $\approx$0.3$c$ at 0.5 mas from the black hole to $\approx$2.7$c$ at $20$ mas (1 mas $\approx 250 \;r_{\rm g}$) \cite{Parketal2019}, showing an efficient conversion of magnetic energy into bulk motion and internal energy of the emitting gas. The kinematic evolution of the jet, however, is quite complex as revealed by numerous investigation \cite{Asadaetal2014,Hadaetal2017,Walker2008}. 

What is the origin of this matter so close to the black hole? In the MAD region the magnetic field is of the order of magnitude given by Equation (\ref{MAD}), i.e., $B_{\rm MAD} \lesssim 10^3$ G~\footnote{For a MAD the ratio $h/r$ is rather small because of the magnetic pressure (say $h/r\approx 0.05$) and then increases with $r$; see the simulations in Ref. \cite{Tchekhovskoyetal2011}. The other values adopted correspond roughly to those of M87.}. This field can efficiently shield the funnel against particle penetration from the disk or its associated wind. The Larmor radius of a~proton with energy $\gamma_p m_p c^2$ is
\beq
r_{\rm L}=\frac{\gamma_p m_p c^2}{eB}\simeq 3\times10^5 \gamma_p \left( \frac{B}{10^3\; \textrm{G}}\right)^{-1} \;\; \textrm{cm}.
\eeq

The Schwarzschild radius $r_{\textrm{S}}= 2r_{\textrm{g}}$ of the SMBH in M87 is of $\sim$10$^{15}$ cm. This sets the scale size of the jet launching region. Clearly, protons, even highly relativistic ones, cannot be directly injected from outside: the magnetic field deflects them on scales that are orders of magnitude smaller than the jet radius. For electrons, the situation is a thousand times worst.  In Sections \ref{pairs} and \ref{baryons} we will discuss some indirect ways to achieve this particle injection at the base of the jet. However, first, it will be useful to have a~better characterization of the material environment of the accreting SMBH in AGNs.

\section{The Black Hole Environment}\label{sec3}

Supermassive black holes lie at the nuclei of galaxies. Their environments often present large amounts of gas and dust that can feed the hole. The sphere of direct gravitational influence of the black hole extends up to the so-called Bondi radius: $r_{\rm B}=2 G M_{\rm BH} / c_{\rm s}^2$ \cite{Bondi1952}, where the gravitational potential energy equals the kinetic energy of the matter. In convenient units:
\begin{equation}
	r_{\rm B}  \simeq 0.1 \left( \frac{k_{\rm B} T}{\rm keV} \right)^{-1} \left( \frac{M_{\rm BH}}{10^9M_\odot } \right)~{\rm kpc} \simeq 10^6~R_{\rm S},
\end{equation}
where $k_{\rm B}$ is the Boltzmann constant and $T$ is the temperature of the gas. The matter that gets into the Bondi sphere falls toward the black hole and can eventually be accreted. The rate of matter crossing the Bondi sphere, $\dot{M}_{\rm B}$, is
\begin{equation}
\dot{M}_{\rm B} = 4\pi r_{\rm B}^2 n m_p c_{\rm s}(r_{\rm B}) \simeq 0.1 \left( \frac{k_{\rm B} T}{\rm keV} \right)^{-3/2} \left( \frac{M_{\rm BH}}{10^9 M_\odot} \right)^2 \left( \frac{n}{0.1~{\rm cm^{-3}}} \right) ~ M_\odot~{\rm yr}^{-1},
\end{equation}
where $n$ is the gas density in the medium, $c_{\rm s}$ is the speed of sound, and $m_p$ is the proton mass. The~definitions given above assume that the matter has negligible angular momentum. This is not the case in most real situations where the gas has non-zero angular momentum and must lose it in order to fall. Assuming that the plasma loses energy faster than angular momentum, matter will drift to the orbit with the lowest energy that is compatible with a given angular momentum, namely to a circular orbit \cite{RomeroVila2014}. In this situation, an accretion disk forms. The physics of these disks is complex and the accretion process develops in many different ways depending on the ambient conditions. To~characterize the different regimes, it is useful to parametrize the accretion rate as a function of the Eddington rate:
\begin{equation}
	\dot{m} = \dot{M} / \dot{M}_{\rm Edd},
\end{equation}
where $\dot{M}_{\rm Edd}=10L_{\rm Edd}/ c^2$ (assuming an efficiency of $10\%$).

At moderate to high accretion rates, $\dot{m} \gtrsim 0.05$, the usual picture of the accretion flow onto supermassive black holes consists of a geometrically thin ($h/r \lesssim 0.1$), optically thick, cold ($T \ll T_{\rm vir}$) disk that cools efficiently, and a geometrically thick ($h/r \gtrsim 0.5$), optically thin, hot ($T \approx T_{\rm vir}$) corona above and below the disk. The corona is thought to be responsible for the hard X-ray tail detected from active galactic nuclei \cite{DiMatteoetal1997,Fabianetal2015}. This emission is produced by hot electrons via inverse Compton up-scattering of UV photons coming from the thin disk.

At lower accretion rates, $\dot{m} < 0.05$, the thin disk does not extend down to the innermost stable circular orbit (ISCO) and it is truncated at longer distances from the hole. The plasma in the inner region is thought to be in the form of a Radiatively Inefficient Accretion Flow (RIAF) \cite{NarayanYi1995, YuanNarayan2014}. A RIAF is~similar to a corona; namely it is a hot, inflated, optically thin plasma. The hot electrons there Compton up-scatter not only photons from the thin disk but also low-energy photons produced by synchrotron radiation. The physical properties of the optically thin hot component in accretion flows, either a corona or a RIAF, are less understood than those of a cold thin disk. In particular, this hot plasma lies in the so-called {\it collisionless regime} of plasmas, and thus particles can be far out of thermal equilibrium. The presence of a non-thermal component leads to the emission of high-energy photons and other particles that might play an important role in the loading of jets, as we discuss in the next~sections.

All this discussion applies to the so-called Standard And Normal Evolution (SANE) state, where the plasma is energetically dominated by the gas and the turbulent magnetic field is subdominant. In~the MAD state, the flow follows a  considerable different evolution in the inner regions. Large-scale magnetic field lines are effectively advected by the accretion flow and accumulate close to the black hole. At a certain distance from the hole, the magnetic pressure balances the action of gravity (see Section \ref{sec:Introduction}) and the flow changes its structure and is accreted through spiralling streams. Despite the fact that in simulations MAD flows show a higher radiative efficiency at equal $\dot{m}$ as compared to SANE disks, the shape of the spectra are quite similar, making it difficult to distinguish the two regimes from observational data \cite{XieZdziarski2019}. 

Hot accretion flows are more efficient than thin disks in launching and collimating relativistic jets. It is not completely clear why this is so, but it is probably related to the more efficient advection of magnetic field lines in hot accretion flows compared to thin disks, and to the much higher geometrical thickness, which might help to confine
and collimate the outflow as well (see Ref. \cite{YuanNarayan2014} for a more extensive discussion). 

The hydrodynamics of a hot accretion flow in an steady state is governed by the following system of equations \cite{YuanNarayan2014}:
\begin{equation}
	\dot{M} = 4\pi r h \rho (-v_r),
\end{equation}
\begin{equation}
	v_r\frac{dv_r}{dr} = \left(\Omega^2 - \Omega_{\rm K}^2 \right)r - \frac{1}{\rho} \frac{d}{dr} \left( \rho c_{\rm s}^2 \right),
\end{equation}
\begin{equation}
	-v_r (\Omega r^2 - j) = \alpha r c_{\rm s}^2,
\end{equation}
\begin{equation}
	q^{{\rm adv},\, i} \equiv \rho v_r \left( \frac{d e_i}{dr} - \frac{p_i}{\rho^2} \frac{d\rho}{dr} \right) = (1-\delta) q^+ - q_{i,e},
\end{equation}
\begin{equation}
	q^{{\rm adv}, \, e} \equiv \rho v_r \left( \frac{d e_e}{dr} - \frac{p_e}{\rho^2} \frac{d\rho}{dr} \right) = \delta q^+ + q_{i, e} - q^-,
\end{equation}
plus an equation of state:

\begin{equation}
	\rho c_{\rm s}^2 = p_{\rm gas} + \frac{B^2}{8\pi} = \left(1+\beta^{-1}\right)p_{\rm gas} = \left(1+\beta^{-1}\right) \frac{\rho k_{\rm B}}{m_p} \left( T_i + T_e \right),
\end{equation}

Here, $\Omega$ and $\Omega_{\rm K}$ are the actual and the Keplerian angular velocity, respectively; $j$ is the angular momentum at the event horizon; $p_i$ ($p_e$) is the pressure of ions (electrons); $e_i$ ($e_e$) is the internal energy of ion (electron) gas; $\alpha$ is the viscosity parameter; $\delta$ is the fraction of energy that directly heats electrons; and $\beta = p_{\rm gas}/p_{\rm mag}$ is the plasma beta parameter. These equations can be solved numerically (see, e.g.,~Refs. \cite{Narayanetal1997, Yuanetal2000}) to obtain the mass density $\rho$, ion and electron temperatures $T_i$ and $T_e$, radial velocity $v_r$, and magnetic field strength $B$.

As a fiducial optimistic example, let us consider a flow around a supermassive black hole with a~mass of $10^9~M_\odot$, extending from $100~R_{\rm S}$ down to the event horizon, and accreting at a rate $\dot{m}=0.01$. Since we are interested in accretion flows in the MAD regime, we adopt $\beta=0.1$, namely we consider a magnetically dominated flow \cite{KimuraToma2020}. The remaining parameters are fixed to standard values, shown in Table \ref{tab:RIAF_parameters}. 

\begin{table}[H]
        \centering
        \caption{Parameters for our fiducial model of a magnetized hot accretion flow.}
        \begin{tabular}{l c}
                \toprule
                \textbf{Parameter [Units]} & \textbf{Value} \\ \cmidrule{1-2}
                $M_{\rm BH}$ black hole mass [$M_\odot$] &  $10^9$ \\
                $\dot{M}$ accretion rate [$\dot{M}_{\rm Edd}$] &  $0.01$ \\
                $r_{\rm out}$ outer radius [$R_{\rm S}$] &  $100$  \\
                $\alpha$ viscosity parameter &  $0.3$  \\
                $\beta$ plasma parameter & $0.1$ \\
                $\delta$ fraction of energy heating electrons &  $0.1$  \\
                \cmidrule{1-2}
                $\xi$ fraction of the accretion power going to non-thermal particles &  $0.05$ \\
                $L_{e,p}$ ratio of electron to proton non-thermal power & $10^{-3}$ \\
                $\eta_{\rm acc}$ acceleration efficiency & $10^{-3}$ \\
                $p$ spectral index of injection & $2$ \\
                \bottomrule
        \end{tabular}
        \label{tab:RIAF_parameters}
  \end{table}
    
\begin{figure}[H]
\centering
\includegraphics[width=0.5\textwidth]{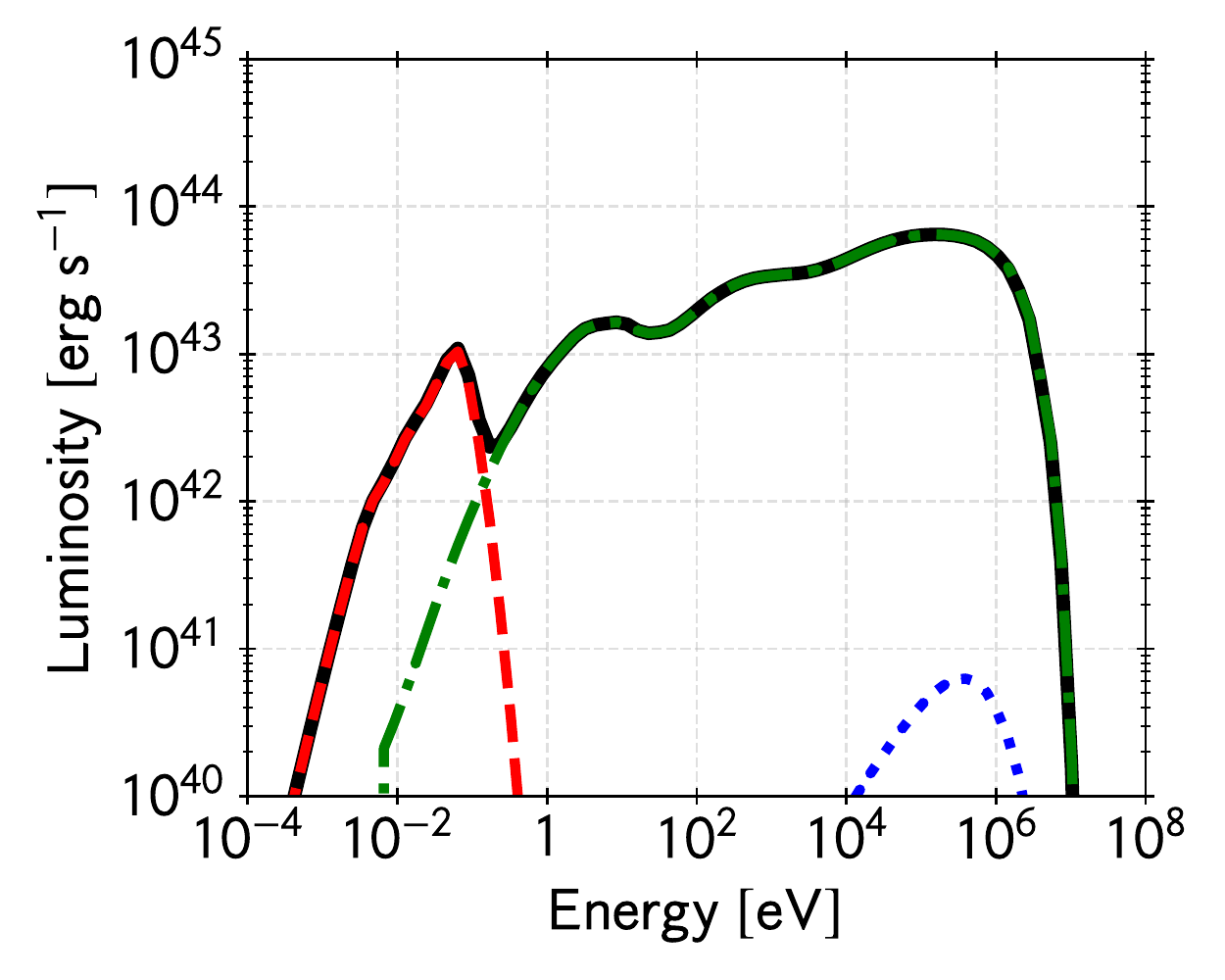}
\caption{Spectral energy distribution for the emission of thermal electrons in the hot accretion flow for our fiducial model (see Table \ref{tab:RIAF_parameters}). The different emission processes are shown. Synchrotron: {\it red dashed}, Bremsstrahlung: {\it blue dotted}, and Inverse Compton: {\it green dot-dashed} (several peaks). The black solid curve is the total emission.}
\label{fig:thermal_SED}
\end{figure}

Thermal electrons reach relativistic temperatures ($T_e \gtrsim 10^9~K$ in the inner region) and cool by synchrotron emission, Bremsstrahlung due to both electron-electron and electron-ion collisions, and multiple Compton up-scatterings of low-energy photons. Figure \ref{fig:thermal_SED} shows the spectral energy distribution (SED) produced by the hot accretion flow defined by the parameters in Table \ref{tab:RIAF_parameters}, taking into account all these processes. Synchrotron emission dominates at millimeter wavelengths\footnote{To facilitate the comparison we remind that the range between 30 GHz and 300 GHz corresponds to a wavelength range of 10 to 1 mm. The photon energy ranges from 0.1 to 1.2 milli-electron volts (meV) in the same interval.}, whereas the inverse Compton radiation completely dominates over Bremsstrahlung at higher energies up to the MeV range.

Let us now consider that a fraction of the particles in the plasma is being steadily accelerated to relativistic energies. Several mechanisms are expected to inject non-thermal particles in hot accretion flows. The most plausible ones are diffusive shock acceleration (DSA) (viable only in regimes that are not magnetically dominated \cite{Drury1983}), magnetic reconnection \cite{Drury2012,Lazarianetal2015}, and stochastic diffusive acceleration (SDA) via turbulence \cite{StawarzPetrosian2008}. 

Without getting into much detail, we assume that a non-specified mechanism acts upon a fraction of the particles in the plasma and accelerates them to a power-law energy distribution of index $\alpha$:
\begin{equation}
	Q(E) = Q_0 E^{-\alpha} {\rm e}^{-E/E_{\rm cut}}, 
\end{equation}
where $Q (E)$ is the injection function of the particles, $E_{\rm cut}$ is the cutoff energy, and $Q_0$ is a normalization constant. We assume that the non-thermal power is $L_{\rm inj} \equiv \int E Q(E) dE = \xi \dot{M}c^2$, where $\xi<1$. The~parameter $\xi$ cannot be very high, otherwise the pressure of locally accelerated relativistic particles, i.e., comsic rays, might exceed the thermal pressure. We choose an optimistic value of $\xi = 0.05$, and the cosmic ray pressure remains $\lesssim 10\%$ of the thermal pressure. An additional observational constrain on this parameter can be imposed by the diffuse neutrino flux (see, e.g., \cite{Kimuraetal2015}). Since our model considers radial dependence, we divide the power into the different regions in the RIAF in such a way that $Q_0(r) \propto u_{\rm th}(r) (|v_r|/r)$, where $u_{\rm th}(r)$ is the energy density in thermal particles at the position $r$. The~cutoff energy is obtained through the balance of acceleration, escape, and cooling timescales. We~parametrize the acceleration time as $t_{\rm acc} = E/\eta_{\rm acc}eBc$, where $\eta_{\rm acc}=10^{-3}$ is the acceleration efficiency and $e$ is the electron charge. To obtain the steady-state particle energy distribution $N(E,r)$, we then solve numerically the following transport equation (e.g., \cite{Schlickeiser2002}):
\begin{equation}
	\frac{1}{r^2} \frac{\partial}{\partial r} \left[ v_r(r) r^2 N(E,r) \right] + \frac{\partial}{\partial E} \left[ b(E,r) N(E,r) \right] + \frac{N(E,r)}{t_{\rm diff}} = Q(E,r).
\end{equation}
Here, $b(E,r) \equiv dE/dt$ is the energy loss rate by all cooling processes considered and $t_{\rm diff}$ is the spatial diffusion timescale. The first term in this equation accounts for the advection of particles, the second for the radiative losses, and the third for the diffusive escape of particles. In steady-state these terms should be equal to the injection function of particles. 

Once we obtain $N(E,r)$, we can proceed to calculate the different processes leading to the radiation of high-energy photons and other by-products. The radiative mechanisms include synchrotron radiation and inverse Compton scattering for electrons, and synchrotron, inelastic proton-proton collisions ($pp$) and proton-photon ($p\gamma$) interactions for protons. In addition, we calculate the production of secondary particles, namely charged pions, muons and secondary electron/positron pairs. Charged pions are produced by inelastic $p\gamma$ and $pp$ collisions, and decay into neutrinos and muons with a~mean lifetime\footnote{Particle lifetimes are always given in the proper system.} of $\tau_{\pi^\pm}\simeq 2.6 \times 10^{-8}$ s. Muons, in turn, decay into neutrinos and electrons/positrons with a mean lifetime of $\tau_{\mu}\simeq 2.2 \times 10^{-6}$ s. The longer lifetime of muons makes them able to cool by synchrotron emission and contribute to the high-energy electromagnetic spectrum \cite{ReynosoRomero2009}. Electron/positron pairs are also produced directly by photo-hadronic collisions (via the Bethe-Heitler channel) and photo-pair production ($\gamma + \gamma \rightarrow e^+ + e^-$). For a detailed account of the model applied here and for a deeper discussion about non-thermal processes in hot accretion flows, the reader is referred to Guti\'errez et al. (2020, in prep.) and Refs. \cite{Romeroetal2010, VieyroRomero2012}.

\begin{figure}[H]
\centering
\includegraphics[width=.67\textwidth]{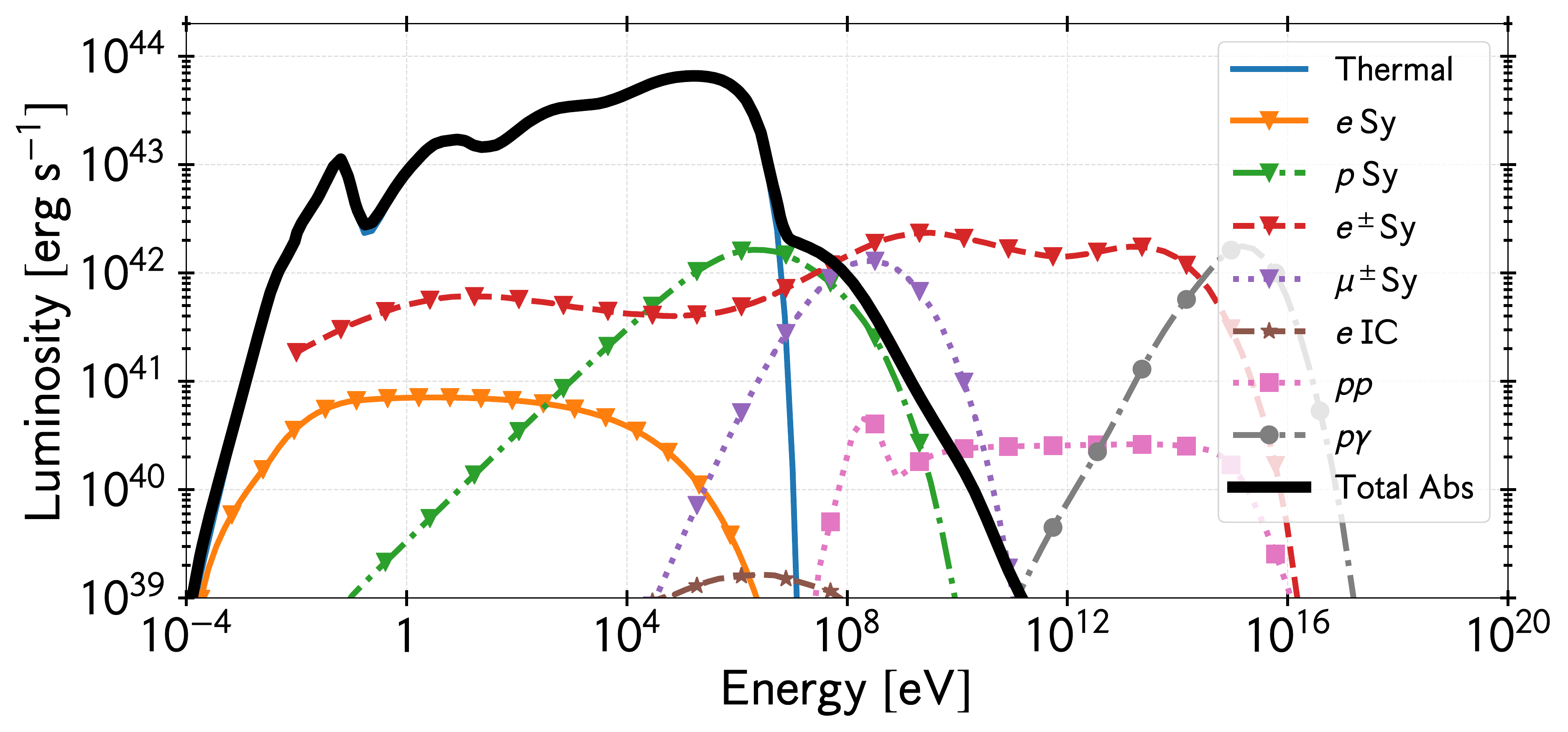}
\caption{Complete spectral energy distribution produced by both thermal and non-thermal particles in the hot accretion flow for our fiducial model (see Table \ref{tab:RIAF_parameters}). The line style, marker, and color of the different emission processes are detailed in the legend. {\it Thermal:} Total emission from thermal electrons (see also Figure \ref{fig:thermal_SED}). {\it $e$ Sy:} Synchrotron emission by primary nonthermal electrons. {\it $p$ Sy:} Synchrotron emission by nonthermal protons. {\it $e^\pm$ Sy:} Synchrotron emission by secondary electrons and positrons. {\it $\mu^\pm$ Sy:} Muon synchrotron emission. {\it $e$ IC:} inverse Compton emission by primary nonthermal electrons. {\it $pp$:} Neutral pion decay via inelastic proton-proton collisions. {\it $p\gamma$:} Neutral pion decay via inelastic photohadronic collisions. {\it Total Abs:} Total electromagnetic radiative output taking into account internal photo-absorption.}
\label{fig:nonthermal_SED}
\end{figure}

Taking into account all processes mentioned, we calculate the non-thermal electromagnetic emission for the model defined by the parameters listed in Table \ref{tab:RIAF_parameters}. Figure \ref{fig:nonthermal_SED} shows the resulting SED with all the processes involved. Synchrotron and Inverse Compton emission from directly accelerated electrons is negligible given the low value of $L_{e,p}$ assumed, but secondary electron/positron pairs produced by the process mentioned above completely dominates the emission in the high-energy band. Proton-proton interactions produce a negligible contribution because of the very low plasma densities. Proton and muon synchrotron radiation contribute in the MeV-GeV range, and pion decay via photo-hadronic interactions ($p\gamma$) is the dominant emission mechanism at very high energies. Despite the high intrinsic gamma-ray luminosities that can be achieved in these systems, very little or no high-energy emission is expected to escape because the compactness parameter\footnote{The dimensionless
compactness parameter $l$ of a gamma-ray source of size $R$ is defined by $l \equiv L \sigma_{\rm T} /Rm_e c^3$, where $L$ is the luminosity, $\sigma_{\rm T}$ is the Thompson cross section and $m_e$ the electron rest mass. A compactness $l<1$ means that the source radiation is self-absorbed through pair creation.} is very large and internal absorption swallows most high-energy photons, with the consequent reprocessing.

Contrarily to gamma rays, neutrinos are not absorbed and can escape. Current observations seem to bound the fraction of the accretion power in RIAF that goes to cosmic ray protons to less than 1\%~\cite{Kimuraetal2015}. Although we have $\xi=0.05$ in our fiducial model (see Table \ref{tab:RIAF_parameters}), we do not intend this model to be representative of typical neutrino sources, which in general might have a lower nonthermal hadronic content \cite{Kimuraetal2019}.

\begin{figure}[H]
\centering
\includegraphics[width=0.36\textwidth]{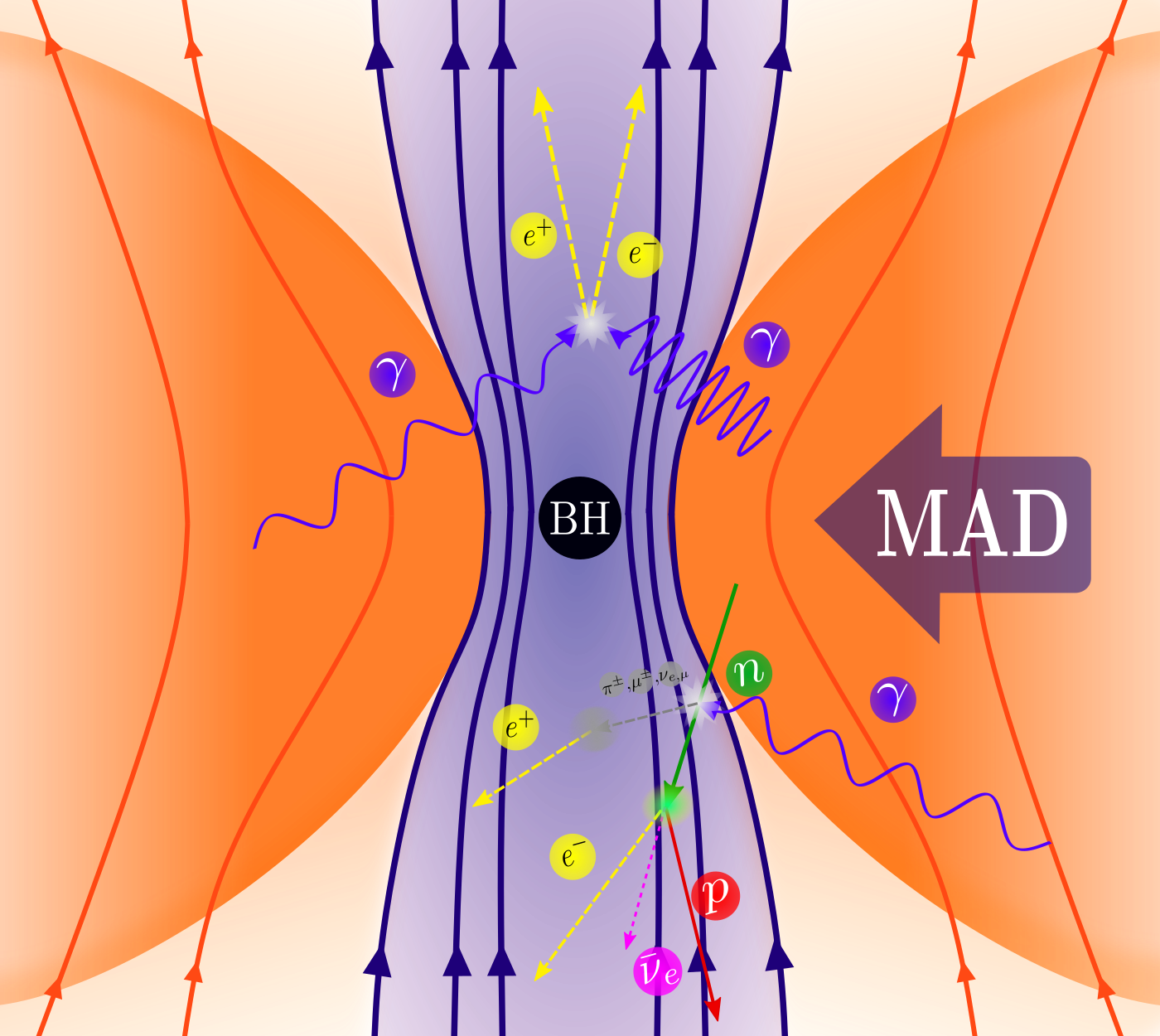}
\caption{Cartoon showing the different components of a MAD flow onto a black hole and the main particle injection mechanisms at the base of the jet: injection of pairs via photon annihilation and injection of baryons (protons) via neutron-driven decays.}
\label{fig:mad}
\end{figure}

In what follows, we shall investigate how neutral particles produced in the immediate surroundings of the black hole through some of the processes discussed above can result in various mass loading mechanisms at the base of the jet. Figure \ref{fig:mad} is a cartoon where the main mechanisms are illustrated. 

\section{Injection of Pairs\label{pairs}}
\unskip
\subsection{Direct Photon Annihilation\label{sec:photon_ann}}

Direct photon annihilation might occur in two ways in the context of hot accretion flows: either two photons with similar energy ($\sim$MeV) collide \cite{Moscibrodzkaetal2011, LevinsonRieger2011}, or a high-energy photon annihilates with a~low-energy photon. Both interactions result in the creation of an electron-positron pair. MeV photons are produced in hot accretion flows by multiple inverse Compton up-scatterings of low-energy photons albeit in low numbers. If non-thermal processes take place in the flow, high-energy gamma-rays are generated and then absorbed in the synchrotron infrared field (see Figure \ref{fig:nonthermal_SED}). 

The opacity to the propagation of a gamma-ray photon of energy $E_{\gamma}$ in a photon field of density $n_{\rm ph}(\epsilon, r)$  is \cite{GouldSchreder1967}
\begin{eqnarray}
	\tau_{\gamma \gamma}(E_{\gamma})=\frac{1}{2}\int_{l}\,\int^{\epsilon_{\rm max}}_{\epsilon_{\rm th}}\int^{\mu_{\rm max}}_{-1} (1-\mu) ~ \sigma_{\gamma \gamma}(E_{\gamma}, \epsilon, \mu)~ n_{\rm ph}(\epsilon, r)\; \mathrm{d}\mu \; \mathrm{d}\epsilon \;\mathrm{d}l.
\end{eqnarray}
Here, $\mu=cos \vartheta$, where $\vartheta$ is the angle between the momenta of the colliding photons, $l$ is the photon path, and the cross section for the interaction is given by
\begin{eqnarray}
	\sigma_{\gamma\gamma}(E_\gamma,\,\epsilon,\vartheta)&=&\frac{3}{16}\sigma_{\rm T}(1-{\beta}_{e^{\pm}}^2)\nonumber\\&&\times\left[\left(3-{\beta}_{e^{\pm}}^4\right)\ln\left(\frac{1+{\beta}_{e^{\pm}}}{1-{\beta}_{e^{\pm}}}\right)-2{\beta}_{e^{\pm}}\left(2-{\beta}_{e^{\pm}}^2\right)\right],
\end{eqnarray}
where ${\beta}_{e^{\pm}}$ is the speed of the electron/positron in the center of momentum frame. This latter quantity is related to the energy of the incident and target photons by
\begin{equation}
(1-{\beta}_{e^{\pm}}^2)=\frac{2m_e^2c^4}{(1-\mu)E_\gamma\epsilon}; \qquad 0\leq{\beta}_{e^{\pm}}<1.	
\end{equation}
The threshold energy $\epsilon_{\rm th}$ is defined by the condition ${\beta}_{e^{\pm}}=1$ with $\vartheta=0$. 

\begin{figure}[H]
\centering
\includegraphics[width=0.75\textwidth]{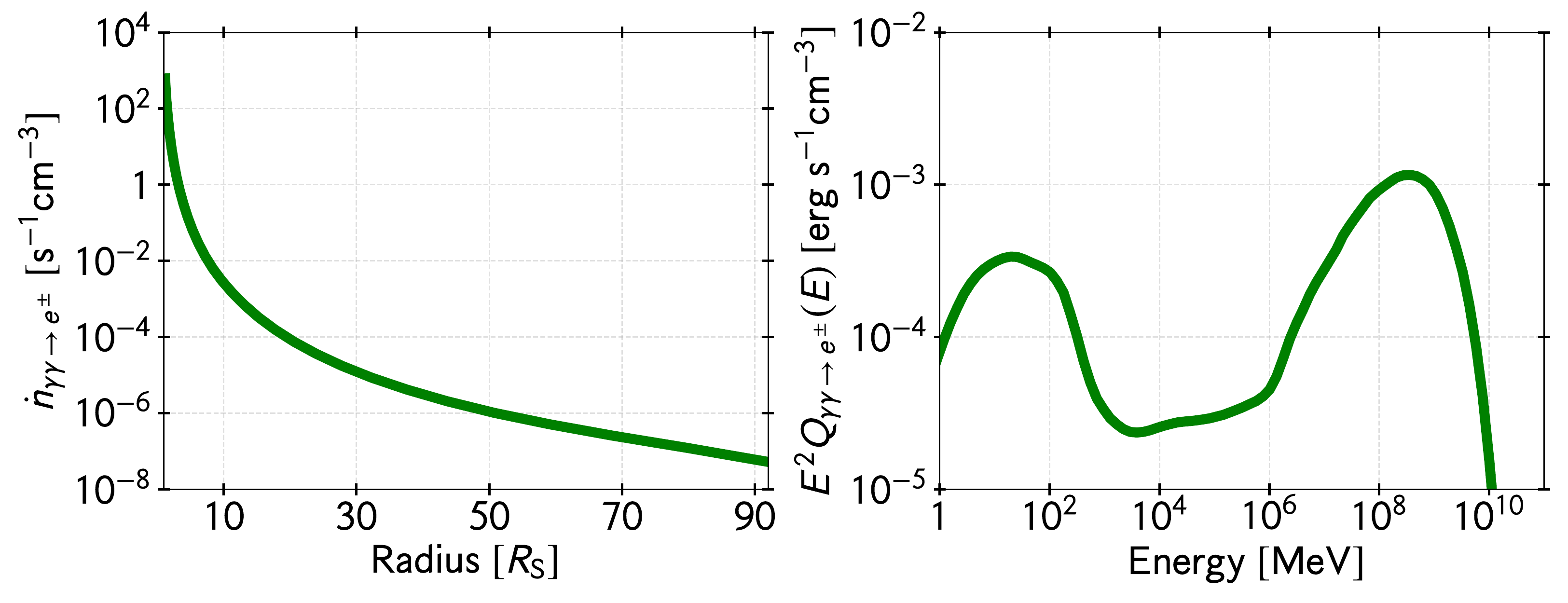}
\caption{Electron-positron pair production in the accretion flow for the fiducial model discussed in the text. {\it Left panel:} Number of pairs created per unit time and volume by photon annihilation in the hot accretion flow as a function of the radius in units of the Schwarzschild radius (this is the distance to the black hole, along the equatorial plane of the disk). {\it Right panel:} Spectral energy distribution of these electron/positron pairs in the innermost region.}
\label{fig:pairs} 
\end{figure}

As can be seen from the $e^{\pm}$ synchrotron curve in Figure \ref{fig:nonthermal_SED}, most of the radiation produced by non-thermal particles above 100 MeV is converted into pairs. 
These pairs are produced at different rates and numbers in different locations of the flow due to the changing conditions. Since photons can penetrate the magnetic barrier of the jet, they also annihilate within the funnel, injecting pairs at the very base of the jet. These pairs are then dragged along with the outflow producing synchrotron radiation. The left panel in Figure \ref{fig:pairs} shows the number of electron-positron pairs created per unit time and volume by photon annihilation within the accreting flow as a function of distance to the black hole. The right panel in Figure \ref{fig:pairs} shows the spectral energy distribution of these secondary $e^{\pm}$ pairs close to the black hole. Two peaks clearly dominate the distribution, one at $E_{e^{\pm}}\sim 70$ MeV and the other at $E_{e^{\pm}}\sim 0.5 $ PeV. The first peak is caused by the absorption of TeV photons in the IR synchrotron radiation of the disk, whereas the second is due to the annihilation of the gamma rays from $p\gamma$ interactions in the IC radiation peak at 100 MeV that can be seen in the SED show in Figure \ref{fig:nonthermal_SED}.

The number density in the innermost region will be comparable to that on the polar funnels. Thus,~since the volume of the base of the jet is $V_{\rm base}\sim 2 \pi R^2_{\rm S} \;h$, with $h$ the height of the funnel (adopted here to be $\sim R_{\rm S}$), the total number of electron/positron pairs resulting from photon annihilation is roughly $N_{e^{\pm}}\approx  \dot{n}_{\gamma\gamma \rightarrow e^{\pm}} V_{\rm base} (R_{S}/c)\sim 10^{55}$. The production rate is $\sim 10^{48}$ s$^{-1}$. Photon annihilation, then, might be a significant lepton loading mechanism as long as non-thermal acceleration processes are active and efficient in the flow around the black hole.  In Section \ref{sec:total} we present the results of more exact numerical estimates based on our fiducial model.

\subsection{Cascades}
\label{subsect:cascades}

If a medium is opaque to radiation and the absorption produces energetic pairs, then a cascade can be generated, in which the number of particles is increased. The cascade develops until the new generations of photons fall below the threshold for pair production.  Electromagnetic cascades in environments with radiation and magnetic fields can develop most efficiently in regimes with very strong fields (i.e., such that $ \epsilon_B/\epsilon_{\rm rad} \gg 1$, with $\epsilon$ the energy density of the magnetic field and the radiation) or if the radiation dominates ($ \epsilon_{\rm rad} \gg \epsilon_B$). In the first case (in magnetized compact objects such as pulsars or magnetars, for instance), high-energy photons are produced by curvature radiation and the absorption is one-photon annihilation in the magnetic field. In the second case, the cascade is the result of the interplay between photon-photon annihilation and IC scattering. 

In hot accretion flows only the second situation occurs, and always outside the MAD region. Hence, cascades are not relevant for mass load at the base of the jet, which is a magnetically dominated zone (i.e., the magnetization parameter is $\sigma \gg 1$). The situation, however, can be different at lower accretion rates. Black hole magnetospheres are thought to behave as a force-free plasma provided a sufficient amount of electron-positron pairs is supplied to them. The pair density required is the so-called Goldreich-Julian density \cite{GoldreichJulian1969}:
\begin{equation}
	n_{\rm GJ} \sim	\frac{\Omega_B B_{\rm z}}{2\pi e c},
\end{equation}
where $B_{\rm z} $ is the component of the magnetic field in the direction of the spin of the black hole. 
When the supply of charges is below this value an unscreened electrostatic field might form, creating a vacuum electrostatic gap in the magnetosphere \cite{BlandfordZnajek1977,Levinson-Segev2017}. For a nearly maximally rotating black hole, the potential drop in the gap is \cite{Levinson2000}
\begin{equation}
	\Delta V \sim 4.5 \times 10^{20}~\left( \frac{M_{\rm BH}}{10^9 M_\odot} \right) \left(\frac{B}{10^4{\rm G}} \right) \left(\frac{h_{\rm gap}}{R_{\rm g}} \right)~~{\rm V},
\end{equation}
where $h_{\rm gap}$ is the gap height. The equation above shows that charged particles can be accelerated up to very high energies by this electric field. A lepton entering into the gap will be quickly accelerated to the maximum energy achievable where the acceleration rate equals the cooling rate. The synchrotron emission affects only the transverse motion of the lepton, whose motion will thus be aligned with the magnetic field lines \cite{BroderickTchekhovskoy2015}. If the leptons emit gamma rays with energies above the photopair-creation threshold, then an electromagnetic cascade might develop and as a result the magnetosphere will be filled with charges \cite{Beskinetal1992, HirotaniOkamoto1998}. Under favourable conditions, the pairs can continue generating cascades outside the gap and the multiplicity can rise $\gtrsim 100$ times \cite{BroderickTchekhovskoy2015}.
When the density of photons supplied by the RIAF is high, the cascades are too efficient within the gap and the electrostatic potential is screened; under this conditions leptons cannot be accelerated anymore \cite{Gutierrezetal2020}. This is the case for our fiducial model. Indeed, Figure \ref{fig:pairs}
 shows that $\dot{n}_{e^\pm} \gtrsim 100~{\rm cm}^{-3}{\rm s}^{-1}$ and hence the density of secondary pairs is $n_{e^\pm} \sim \dot{n}_{e^\pm} \times (R_{\rm S}/c) \gtrsim 10^6~{\rm cm}^{-3}$. Comparing this number to the Goldreich-Julian density $n_{\rm GJ} \sim \Omega_B B_{\rm n}/2\pi e c \sim 5 \times 10^{-3}~{\rm cm^{-3}}$, we obtain $n_{e^\pm} \gg n_{\rm GJ}$. Thus, any electrostatic potential would be quickly screened.

\subsection{Bethe-Heitler Mechanism}

Proton-photon ($p\gamma$) inelastic collisions occur via two main channels: photo-pair production (Bethe-Heitler effect) and photo-meson production. The first channel, 
\begin{equation}
p+\gamma\rightarrow p+e^-+e^+,
	\label{photopair}
\end{equation}  
\noindent with a threshold energy of $2m_e c^2 \sim 1$ MeV for the photon in the hadron rest frame, can be particularly important for particle injection in the jet. Neutrons produced in the hot flow can escape and move through the funnel of the jet without being affected by the magnetic field. There, they can decay, transferring most of their energy into the protons produced which, in turn, will generate numerous pairs in the IR background photon field. This process will be important only if a non-thermal population of protons in the accretion flow produces the neutrons since high Lorentz factors are required for the latter to reach the funnel\footnote{ The mean lifetime of neutrons in the disk reference system is $\tau_n\simeq 880 \gamma_n$ s. Since $R_{\rm S}\sim 3 \times 10^{14}$ cm, neutrons should have Lorentz factors $\gamma_n\gtrsim 10$, i.e., energies above $\sim$10 GeV.}. 

The $p\gamma$ cooling rate by Bethe-Heitler interactions for a proton of Lorentz factor $\gamma_p$ in a photon field of density $n\left(\epsilon\right)$ is \cite{Begelmanetal1990,Masrtichiadisetal2005,RomeroVila2008} 
\begin{eqnarray}
t^{-1}_{p\gamma, \,e^\pm}(\gamma_p)=\frac{c}{2\gamma_p^2}&& \int_{\epsilon_{\rm{th}}/2\gamma_p}^\infty \mathrm{d}\epsilon  \frac{n\left(\epsilon\right)}{\epsilon^2}\nonumber\\&& \times\int_{\epsilon_{\rm{th}}}^{2\epsilon\gamma_p}\mathrm{d}\epsilon^\prime\sigma_{p\gamma,\,e^\pm}\left(\epsilon^\prime\right)K_{p\gamma,\,e^\pm}\left(\epsilon^\prime\right)\epsilon^\prime,
	\label{coolng}	
\end{eqnarray} 
 
\noindent where  $\epsilon^\prime$ is the energy of the photon in the rest frame of the proton and $\epsilon_{\rm{th}}$ is the photon threshold energy measured in the same frame and $K_{p\gamma, e^\pm}$ is the inelasticity of the process. The corresponding collision rate ($\omega_{p\gamma,\,e^{\pm}}$) is given by a similar expression:  

\begin{equation}
\omega_{p\gamma,\,e^{\pm}}(\gamma_p)=\frac{c}{2\gamma_n^2}\int_{\frac{\epsilon_{\rm{th}}}{2\gamma_n}}^\infty\mathrm{d}\epsilon\frac{n\left(\epsilon\right)}{\epsilon^2}\int_{\epsilon_{\rm{th}}}^{2\epsilon\gamma_p}\mathrm{d}\epsilon^\prime\sigma_{p\gamma,\,e^\pm}\left(\epsilon^\prime\right)\epsilon^\prime.
	\label{Kpg}	
\end{equation} 
 
The inelasticity can be approximated by its value at the threshold, $K_{p\gamma,\,e^\pm}=2m_e/m_p$, with $m_e$ and $m_p$ being the masses of the electron and the proton, respectively . Therefore,  the injection function $Q_{e^{\pm}}$~results:

\begin{eqnarray}
Q_{e^{\pm}}\left(E_{e^{\pm}}\right)&=&2\int\mathrm{d}E_p N_p\left(E_p\right)\omega_{p\gamma,\,e^\pm}\left(E_p\right)\delta\left(E_{e^{\pm}}-\frac{m_e}{m_p}E_p\right)\nonumber\\
&=&2\frac{m_p}{m_e}N_p\left(\frac{m_p}{m_e}E_{e^{\pm}}\right)\omega_{\gamma,\,e^\pm}\left(\frac{m_p}{m_e}E_{e^{\pm}}\right),
	\label{emispairs}	
\end{eqnarray}  
 A useful parametrization of the cross section $\sigma_{p\gamma,\,e^\pm}$ can be found in Ref. \cite{Maximon1968} (see also Ref. \cite{Chodorowskietal1992}).
 
To produce pairs against the IR peak of the synchrotron radiation produced in the accreting flow (see Figure \ref{fig:thermal_SED}) protons injected through the decay of neutrons penetrating into the funnel should have a Lorentz factor $\gamma_n\sim 10^7$, i.e., energies of $\sim$10 PeV. The neutrons should be generated by very energetic protons in the hot flow through $pp$ or $p\gamma$ collisions. Mildly relativistic protons can also produce pairs or even pions interacting with X-ray or MeV photons corresponding to the high-energy sector of the thermal SED.

A quantitative estimate of the pairs injected by Bethe-Heitler mechanism into the jet requires first an estimate of the neutron luminosity of the hot accretion flow. We turn to this issue now.
 
\section{Injection of Baryons}\label{baryons}

We shall calculate the number of baryons injected in the jet by different processes as a function of height. We shall assume homogeneous jet cross-sections and hence we simply calculate the number of particles injected per unit time per unit volume along the $z$-axis.
\subsection{Neutrons}

\begin{figure}[H]
\centering
\includegraphics[width=0.5\textwidth]{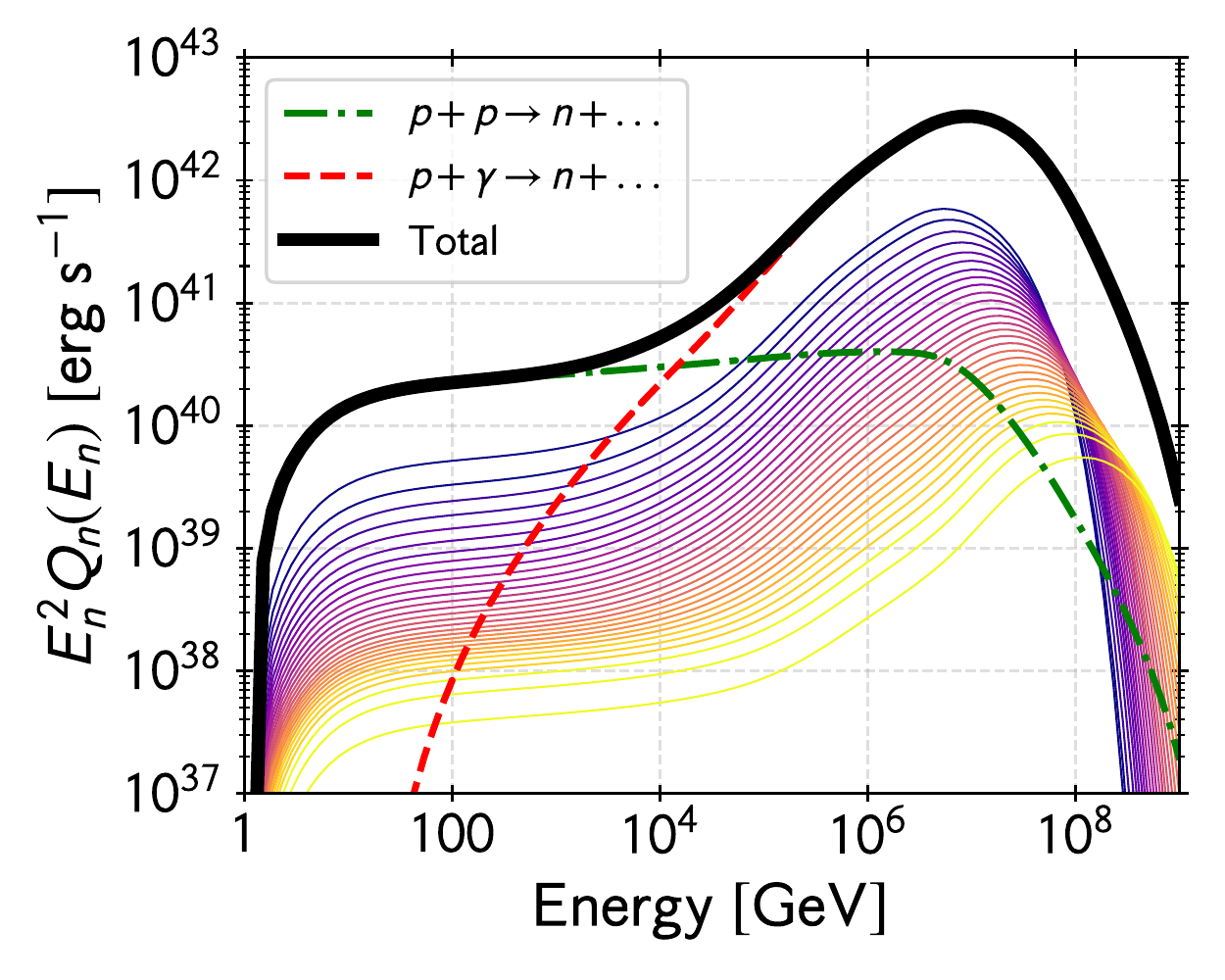}
\caption{Energy distribution of neutron injection in the accretion flow for the fiducial model discussed in the text. The thin lines show the contribution from different parts of the flow; the darker blue curves are located closer to the black hole than the lighter yellow curves. The total contribution summing all regions in the flow is shown as {\it dashed red lines} ($p\gamma$) and {\it green dot-dashed lines} ($pp$). The {\it dark solid} line is~the total injection function.}
\label{fig:neutron_inj}
\end{figure}

\begin{figure}[H]
\centering
\includegraphics[width=0.5\textwidth]{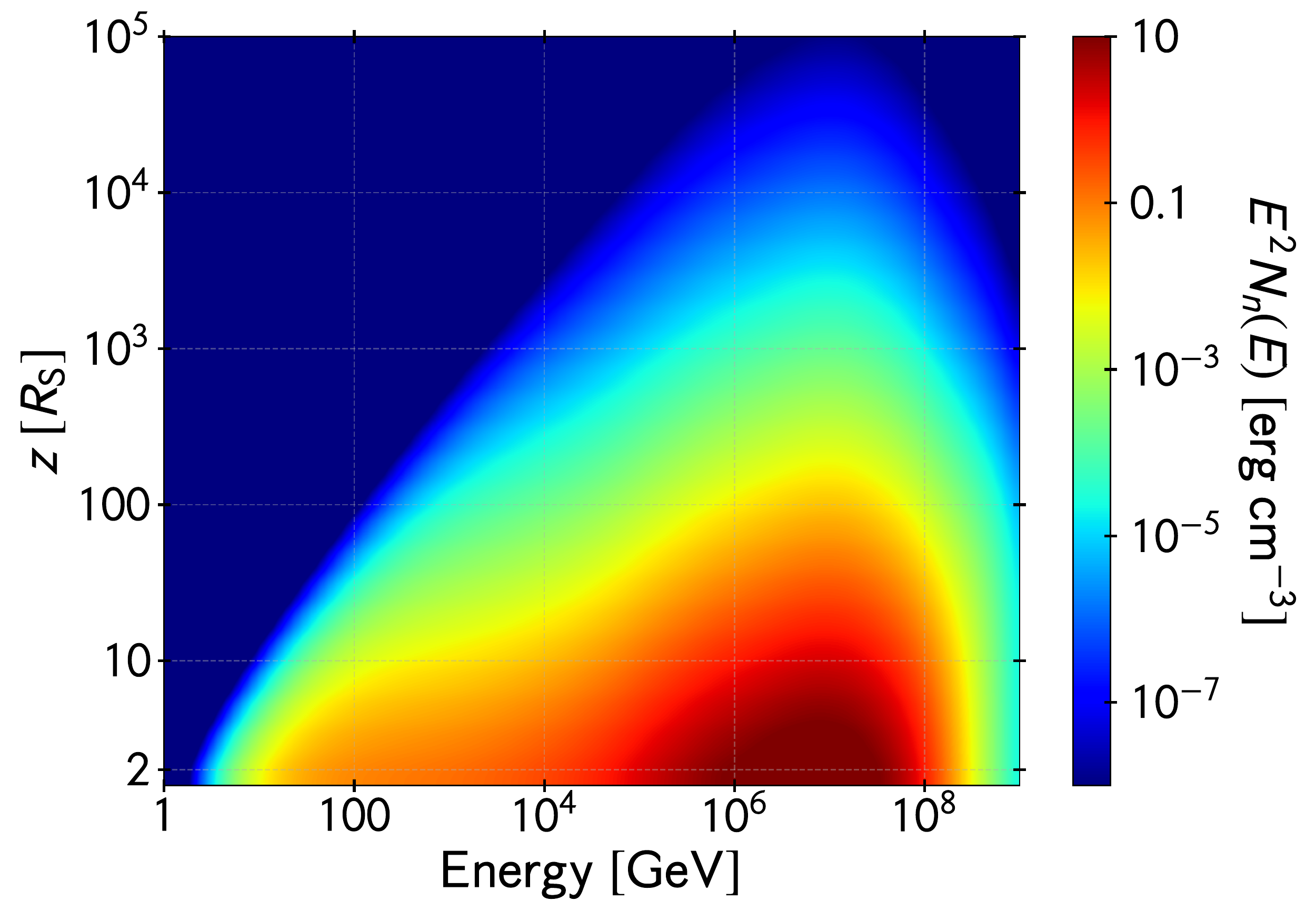}
\caption{Steady neutron energy distribution evaluated on the jet axis. The vertical axis is the distance to the hole in units of Schwarzschild radii and the horizontal axis is the neutron energy. The color map represents the neutron energy density per unit logarithmic bin of energy.}
\label{fig:neutron_dist}
\end{figure}

The relativistic protons in the accretion flow produce neutrons through the reactions 
\begin{eqnarray}
p + p &\rightarrow & p + n + \pi^+ + a\pi^0 + b(\pi^+ + \pi^-), \\
p + p &\rightarrow &  n + n + 2\pi^+ + a\pi^0 + b(\pi^+ + \pi^-),  \\
p + \gamma &\rightarrow &   n + \pi^+ + a\pi^0 + b(\pi^+ + \pi^-), \label{pgamma}\end{eqnarray}

\noindent where $a$, $b$ are integers. Some of these neutrons can escape, penetrate into the base of the jet and decay there, injecting there protons and electrons according to $n \rightarrow p + e^- + \bar{\nu}_e$ \cite{Toma2012, Vilaetal2014}. To estimate these contributions, we adopt the neutron injection functions given by Refs. \cite{Sikoraetal1989,Atoyan1992,AtoyanDermer2003}. Figure \ref{fig:neutron_inj} shows the spectral power injected into neutrons by $pp$ and $p\gamma$ interactions. The thin curves show the contribution from different regions of the flow, being the highest in the innermost region. For our fiducial model, the dominant process of neutron injection comes from photo-meson production. Since neutrons escape from the disk and then decay without significant energy losses, we can estimate the steady-state distribution of neutrons in the funnel of the jet as $N_n(E_n,z) \simeq (4\pi c)^{-1} \int dV Q_n(E_n,r) d_{rz}^{-2} \exp (-d_{rz}/r_\tau )$. Here $d_{rz}=\sqrt{r^2+z^2}$ is the distance from the region of the disk where neutrons are produced and the vertical axis at a height $z$, $r_\tau = \tau_n v_n$, and $\tau(E_n)=\gamma_p \tau_n$ and $v_n(E_n) \approx c$ are the mean lifetime in the lab frame and the velocity field of the neutrons, respectively. Figure \ref{fig:neutron_dist} shows the energy distribution of neutrons on the jet axis. When a neutron decays, the proton takes $\sim$99.9\% of the neutron energy, so we can approximate the proton injection function as $Q^{n\rightarrow p}(E_p,r) \approx N_n(E_n, r)/\tau(E_n)$ with $E_p = 0.999 E_n$. Regarding the electrons, their injection rate $Q_e^{n\rightarrow e^-}$ can be calculated using the parametrization given in Ref. \cite{Abrahametal1966}. Changes in the transverse direction to the jet axis are not significant because of the relatively small dimensions at such close distances to the black hole.

\unskip

\begin{figure}[H]
\centering
\includegraphics[width=0.45\textwidth]{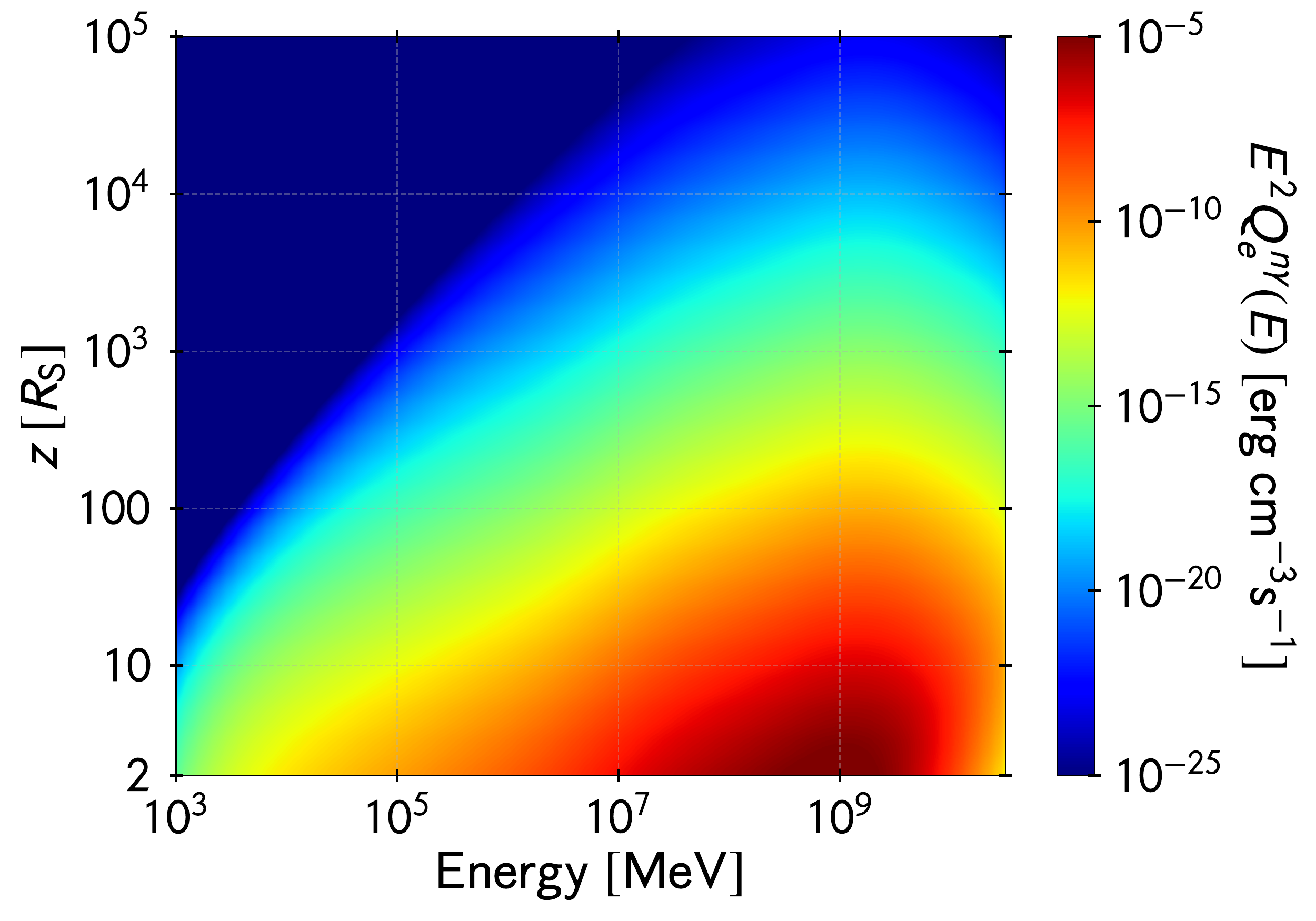}
\caption{Electron injection in the jet axis via photo-meson production by neutrons in the funnel. The~vertical axis is the distance to the hole in units of Schwarzschild radii and the horizontal axis is the energy of the created electron. The color map represents the energy density injected in electrons per unit logarithmic bin of electron energy per unit time.}
\label{fig:e_dist}
\end{figure}

\begin{figure}[H]
\centering
\includegraphics[width=0.45\textwidth]{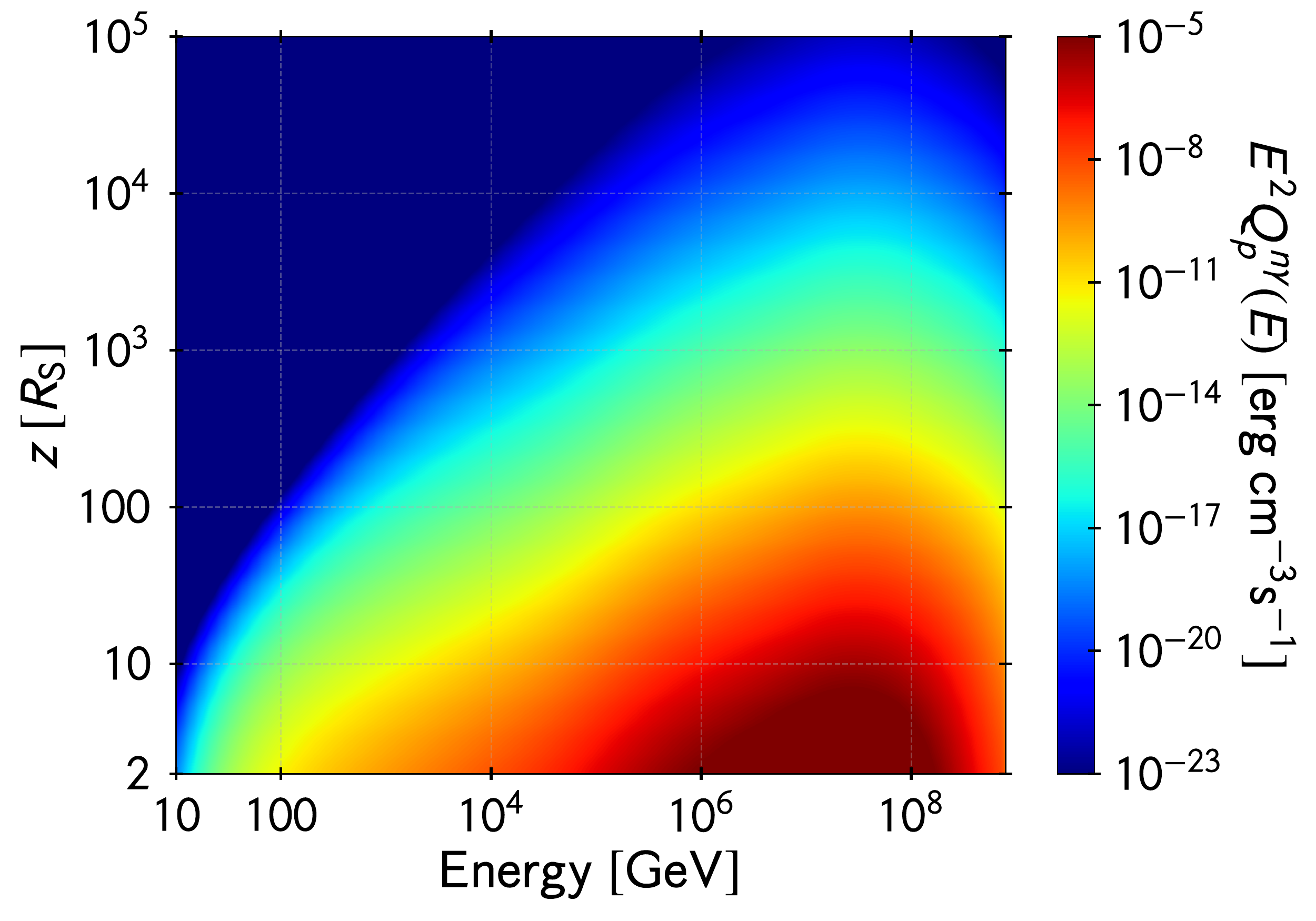}
\caption{Proton injection in the jet axis via photo-hadronic collisions of neutrons in the funnel. The~vertical axis is the distance to the hole in units of Schwarzschild radii and the horizontal axis is the energy of the created proton. The color map represents the energy density in protons injected per unit logarithmic bin of proton energy per unit time.}
\label{fig:p_dist}
\end{figure}

Direct neutron-photon scattering is also possible \cite{Gould1993}. Though the neutron as a whole is electrically neutral, it is not a point object. That is, it is made up of three quarks ($udd$), and the center of negative charge does not coincide with the center of the positive charge. The result is that the neutron has higher multipole moments. A photon can interact with a particle if it has a multipole moment, even~if its net charge is zero. When a photon hits a neutron, it usually gets scattered. There is, however, a~finite chance of the photon getting absorbed and converted into an $e^\pm$ pair. The probability of such a process is very low and no significant contribution to the particle content of the jet should be expected from this phenomenon. On the other hand, neutron-photon collisions can lead to the production of charged pions, which in turn decay into muons and neutrinos. Muons are also transient and decay into an electron or positron and two neutrinos. The main channel for this photo-hadronic interaction is $n+\gamma\rightarrow p + \pi^-$, with the channel $n + \gamma \rightarrow \pi^0$ being negligible \cite{Atoyan1992}. The cross section can be approximated as $\sigma_{n\gamma}\approx \sigma_{p\gamma}$, and the injection of electrons is $E_e^2Q_e^{n\gamma}(Ee) \approx \frac{1}{8}E_n^2N_n(E_n) t_{n\gamma}^{-1}$, where $E_e \approx 0.05 E_n$. These contributions are shown in Figures \ref{fig:e_dist} and \ref{fig:p_dist}. 

\unskip

\begin{figure}[H]
\centering
\includegraphics[width=0.45\textwidth]{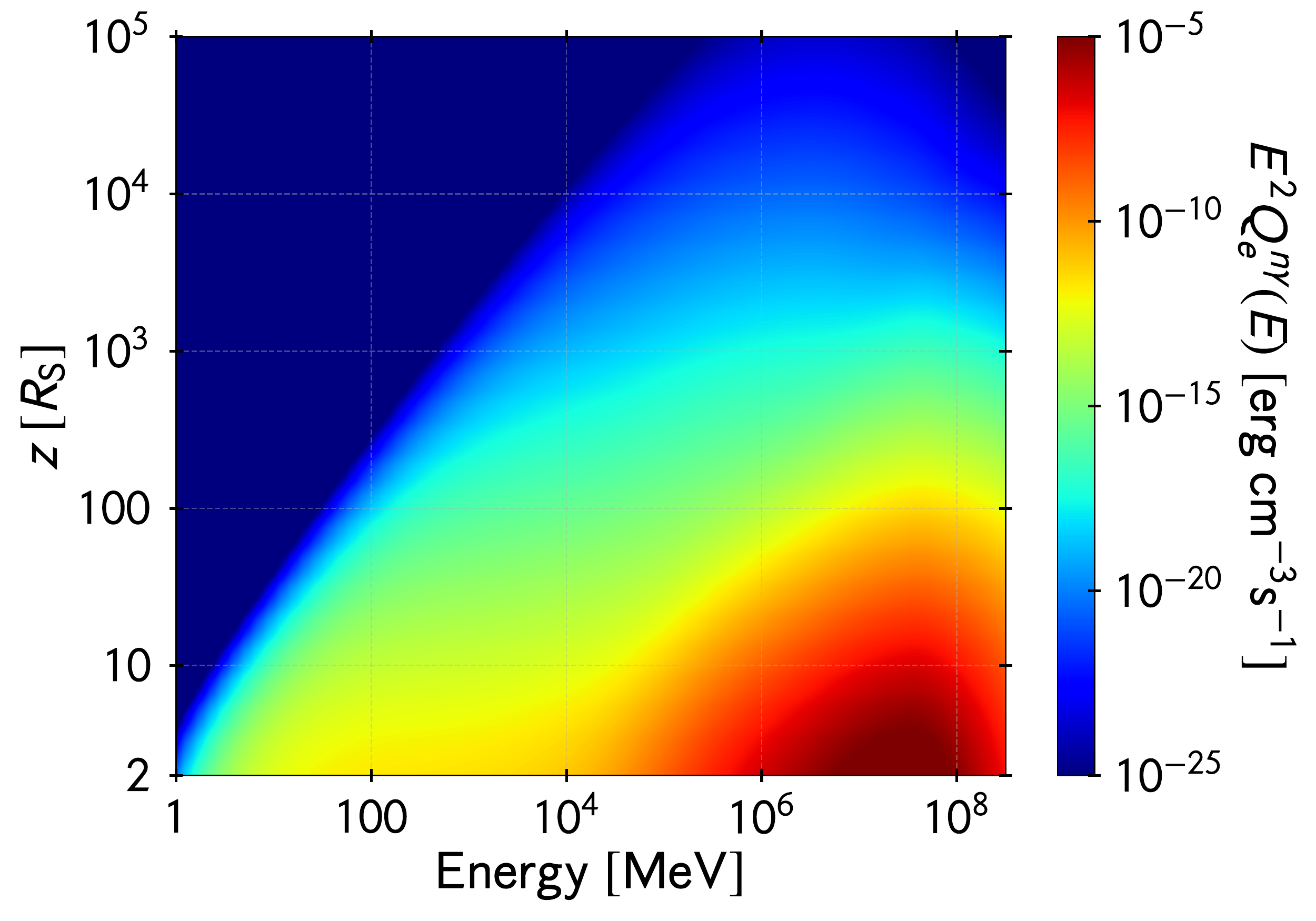}
\caption{Electron injection in the jet axis via the Bethe-Heitler channel by photohadronic collisions of protons in the funnel. The vertical axis is the distance to the hole in units of Schwarzschild radii and the horizontal axis is the energy of the created electron. The color map represents the energy density injected in electrons per unit logarithmic bin of electron energy per unit time.}
\label{fig:e_BH}
\end{figure}

The protons, once injected from $n+\gamma\rightarrow p + \pi^-$ will produce additional pairs by Bethe-Heitler mechanism and, if they are energetic enough, more pions through interacting with the photons that fill the funnel. This secondary pair injection is shown in Figure \ref{fig:e_BH} , for the system characterized by the parameters of Table \ref{tab:RIAF_parameters}.

\subsection{Entrainment}

The funnel that contains and collimates the initial jet is formed by the inflated hot disk and the associated wind. The development of magnetohydrodynamical (MHD) instabilities on the interface between the jet and the external medium might in principle lead to some entrainment of matter in the outflow. Kelvin-Helmholtz instabilities, however, are inhibited by the large dominant poloidal fields~\cite{Romero1995}.  On the other hand, pressure-driven instabilities develop only in the subsonic regime \cite{Kersale2000}. Current driven instabilities are the most relevant ones for Poynting-flux dominated jets, being the mode $m = 1$ kink instability the most effective. The growth of this instability requires the toroidal component of the magnetic field to be important. Such a situation occurs beyond the Alfv\'en radius, i.e., where the magnetic energy density is surpassed by the gas energy density. This happens far beyond the funnel \cite{GianniosSpruit2006}. 

Large-scale fully three-dimensional GRMHD simulations of rapidly rotating, accreting black holes producing jets have shown that the accretion of non-dipolar magnetic fields resulting from turbulence in the accretion disk leads to weak, turbulent outflows, with significant matter loading \cite{McKinneyBlandford2009}. The jet becomes non-relativistic or weakly relativistic in these cases. On the contrary, accretion of dipolar fields generates highly relativistic and stable jets. These results seem to suggest that a fast increase of the matter content of the jet might occur under the right accretion conditions without fully destabilizing the outflow. Most of the entrainment, however, is expected to happen beyond the funnel \cite{Mignoneetal2010}. 

\subsection{Bulk Matter}

When a jet meets an obstacle, such as a star or a cloud, it can be loaded with matter from the stellar wind, the stellar atmosphere, or the ablation of the cloud \cite{BlandfordKoenigl1979,Komissarov1994,Araudoetal2010,BoschRamonetal2012,Araudoetal2013}. These interactions, however, occur far away from the base of the jet, in a region where the flow is matter-dominated. Even at distances $\sim$10$^2$--10$^3$ $r_{\rm g}$, where the interaction is with clouds of the Broad Line Region (BLR), the jet has a magnetization $\varsigma \ll 1$ \cite{delPalacio2019}. 

In the magnetically dominated region, close to the funnel, clumps of gas might exist from turbulence and instabilities happening in the wind produced by the hot disk. The existence of strong winds in hot accretion flows has been established by HD and MHD simulations (e.g., Refs. \cite{Yuanetal2012,YuanNarayan2014,Buetal2016}). In these winds, magnetic turbulence and radiative instabilities might generate a clumped structure as in stellar winds of massive stars \cite{RunacresOwocki2002}. Some of the cloudlets can be impelled in the direction of the jet. The physics of the interaction of a plasma cloud with a jet is complex and not well understood in a magnetically dominated case, but on a very first approximation the penetration of the cloud into the magnetic jet requires at least the magnetic energy density at the base of the jet to be smaller than the cloud kinetic energy $\rho_{\rm cloud} v_{\rm cloud}^2/2>B_{\rm j}^2/8\pi$. For the model specified in Table \ref{tab:RIAF_parameters}, $B_{\rm j}\sim 10^4$ G and we adopt a large transverse velocity of $v_{\rm cloud}\sim 10^8$ cm s$^{-1}$. Then, $\rho_{\rm cloud}\gtrsim 10^{-7}$ g cm$^{-3}$. Simulations suggest much smaller values for the density at the base of the wind \cite{Buetal2018}. We conclude that direct injection of matter from the clumpy medium that surrounds the base of the jet is unlikely.

\section{Total Injection}
\label{sec:total}

\begin{figure}[H]
\centering
\includegraphics[width=.67\textwidth]{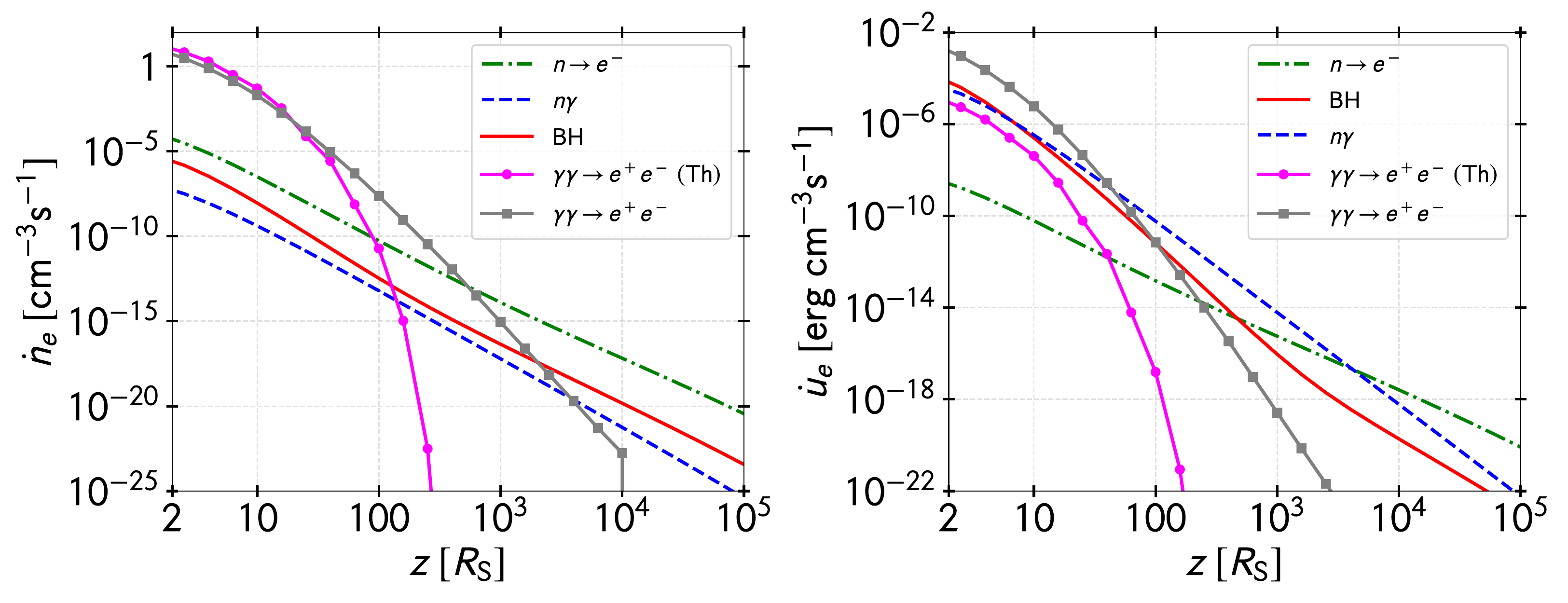}
\caption{Injection of electrons and positrons in the funnel by all processes mentioned in the text. BH stands for Bethe-Heitler. {\it Left panel:} Number of pairs created per unit time and volume by the different processes. The curves of pairs created by photo-annihilation are divided into the contributions from MeV-MeV collisions (magenta line with circles) and those of non-thermal gamma-rays with soft photons (grey line with squares). {\it Right panel:} Same plot for the energy injected in pairs per unit time and volume.}
\label{fig:electron_dens} 
\end{figure}

In this section, we present the combined effect of all different injection channels of massive particles in the funnel of the jet for our model. The left panel in Figure \ref{fig:electron_dens} shows the number of electron-positron pairs created per unit time and volume at the funnel by all processes discussed above. The curves of pairs created by photon-annihilation are divided into the contribution from MeV-MeV collisions (photons mainly produced by thermal particles) and that of the collisions between non-thermal gamma rays and low-energy thermal photons (see Section \ref{sec:photon_ann}). The right panel of the figure shows a similar plot but for the total power injected per unit volume. It can be noticed that the relative importance of the different processes changes with the distance to the black hole.

\begin{figure}[H]
\centering
\includegraphics[width=.67\textwidth]{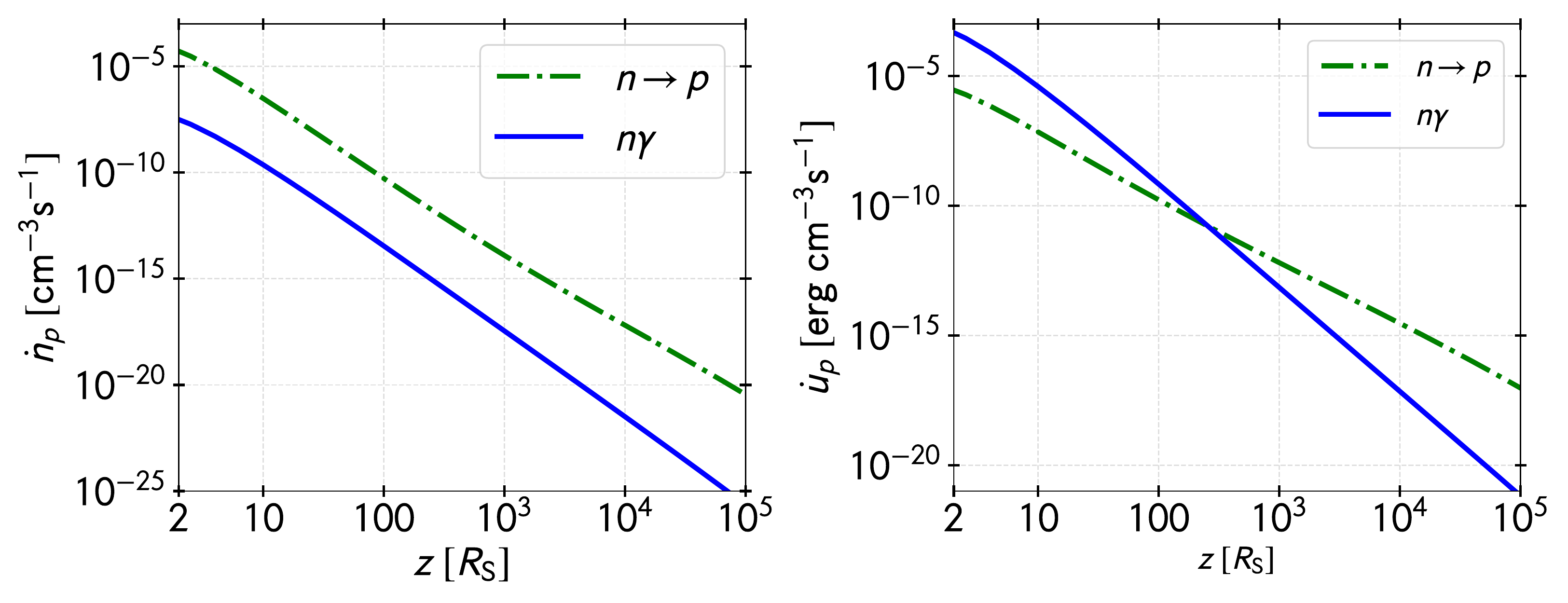}
\caption{Injection of protons in the funnel by the two processes mentioned in the text. {\it Left panel:} Number of protons created per unit time and volume. {\it Right panel:} Same plot for the energy injected in protons per unit time and volume.}
\label{fig:proton_dens}
\end{figure}

Figure \ref{fig:proton_dens} is a similar plot but for protons.

\unskip

From Figures \ref{fig:electron_dens} and \ref{fig:proton_dens} we can see that the dominant mechanism of pair injection close to the black hole is photon annihilation. Thermal and non-thermal annihilations are similar up to less than 100 $R_{\rm S}$ from the black hole. Beyond that non-thermal gamma rays annihilating against soft IR photons become more important. At distances larger than 1000 $R_{\rm S}$ electrons are mainly injected by the decay of very energetic neutrons. From the energetic point of view, however, the non-thermal processes are able to dominate the pair injection, first by photon annihilation and then, beyond 100 $R_{\rm S}$, by meson decay and finally at large distances by neutron decay.

Most protons injected are the result of neutron decay, although direct neutron conversion into protons ($n\gamma\rightarrow p + \pi^-$) might be important close to the black hole. At the very base of the jet, the ratio of the number of pairs to the number of protons is $R_{e^\pm / p}\sim 10^5$ and the associated energy ratio is  $R_{u_e^\pm / u_p}\sim 10^2$. Most of the jet mass load is, then, in the form of pairs.

\subsection*{Magnetization of the Jet}

Jets are launched as highly magnetized outflows \cite{Potter2017}. There are different parameters used to quantify the dynamical importance of magnetic fields in a plasma (see, e.g., Ref. \cite{Komissarov1999}). The most extensively used one is the plasma $\beta$ parameter, defined as the ratio of gas pressure to magnetic pressure: $\beta \equiv p_{\rm gas}/p_{\rm mag}$. However, in high-velocity outflows it is most common to adopt the so-called {\it magnetization} parameter $\sigma$. It is defined simply as the ratio of magnetic energy density to rest mass energy density, though other definitions are also used\footnote{There are several slightly different definitions of this parameter. The most general one takes into account the bulk kinetic energy and the internal energy of the matter: $\sigma \equiv U_B/ (\Gamma_{\rm bulk}\rho c^2 + U_{\rm int})$.}:
\begin{equation}
\sigma = \frac{B^2}{8\pi\rho c^2}.
\end{equation}
In this case, $\sigma$ gives an upper bound for the magnetization in the jet. To calculate the evolution of the magnetization along the jet we must assume a jet shape and a dependence of the magnetic field with the distance on the vertical axis. We parameterize these dependencies as
\begin{equation}
	B(z) = B_0~\left( \frac{z_0}{z} \right)^m,~~~~~~~~~~~R(z) = R_0 \left( \frac{z}{z_0} \right)^b.
\end{equation}
Here, $z$ is the distance along the vertical (jet) axis and $R$ is the radius of the jet cross section. The~index $m \geq 1$ depends on the topology of the magnetic field lines \cite{Krolik1999}. For a conical jet and a purely poloidal magnetic field, the index is $m=2$, whereas for a purely toroidal magnetic field we have $m=1$. Regarding the shape of the jet, $b=0$ corresponds to a cylindrical jet, $b=1/2$ to a parabolic jet, and $b=1$ to a conical jet.

To calculate the matter density, we divide the jet in $N$ cells, so in the $j$-cell the particle density can be estimated as
\begin{equation}
	n_j \approx \frac{\dot{n}_j V_j}{\pi R^2_{j-1}\beta_{\rm bulk}c} + n_{j-1} \left( \frac{R_{j-1}}{R_j} \right)^2,
\end{equation}
where
\begin{equation}
	V_j = \frac{\pi R_0^2 z_0}{2b+1} \left[ \left( \frac{z_j}{z_0} \right)^{2b+1} - \left( \frac{z_{j-1}}{z_0} \right)^{2b+1} \right]
\end{equation}
is the volume of the $j$-cell and $z_j$, $R_j$ are the height at the boundary between the $j$-cell and the $(j+1)$-cell ant its cross-section radius.

\begin{figure}[H]
\centering
\includegraphics[width=0.45\textwidth]{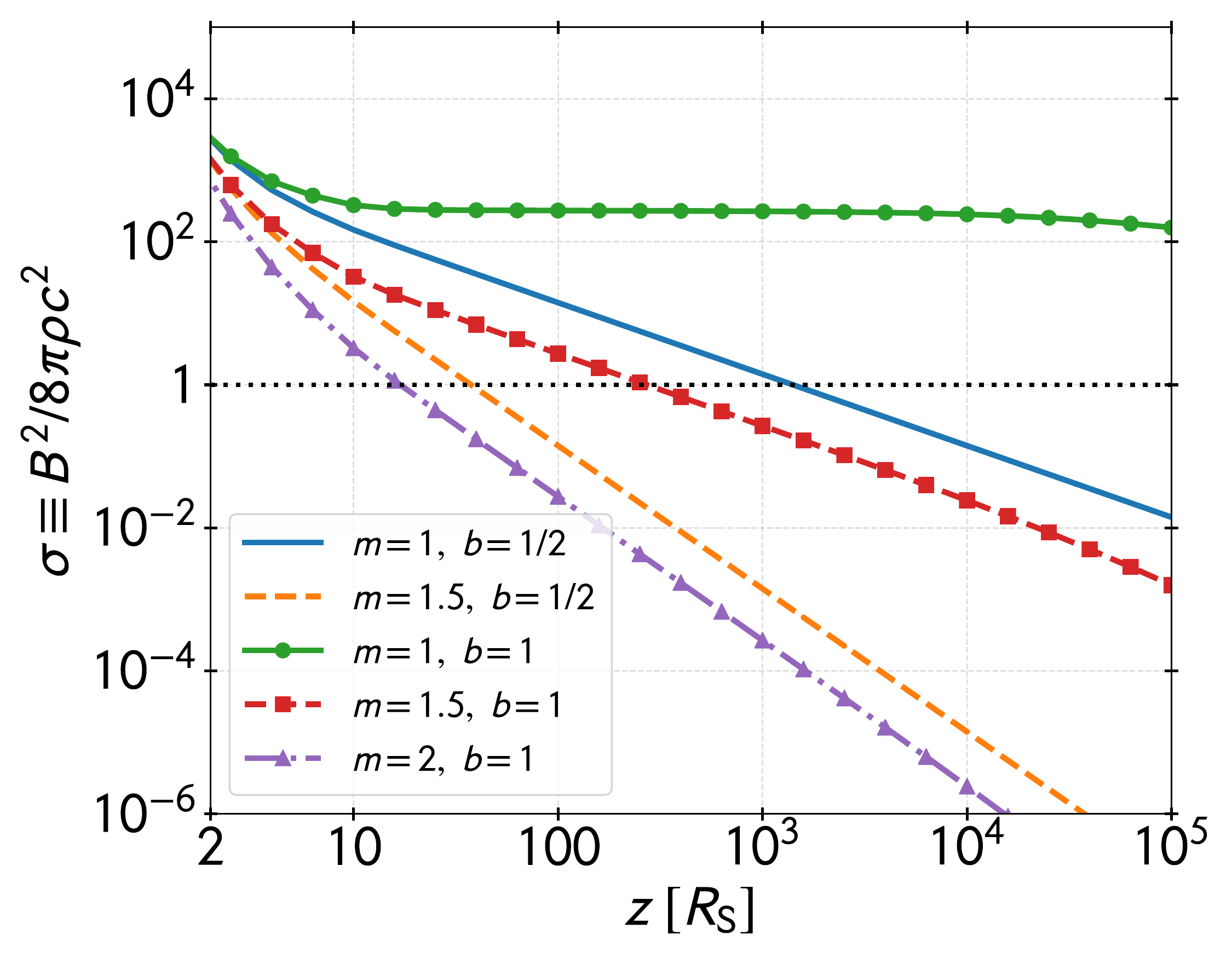}
\caption{ Magnetization along the jet for different topologies of the magnetic field and different jet shapes. The cases where $b=1$ correspond to a conical jet and $b=1/2$ to a parabolic jet. The magnetic field intensity decays as  $\sim$z$^{-m}$, where $m = 1$ represents a purely toroidal magnetic field and $m = 2$ a purely poloidal field.} 
\label{fig:magnetization}
\end{figure}

Figure \ref{fig:magnetization} shows the magnetization parameter as a function of the distance to the black hole along the jet. We have taken $z_0=R_0 \approx R_{\rm S}$ and a magnetic field intensity at the base $B_0=10^4~{\rm G}$. We show different curves for different choices of the magnetic field topology and jet shape. Likely, the jet has a parabolic shape at the base due to the confinement imposed by the accretion flow. The magnetic field topology is less clear, though the magnetic field is certainly highly poloidal close to the launching point and toroidal far from the hole. For a parabolic shape ($b=1/2$) and a mixed magnetic topology ($m\sim 1.5$), the jet becomes matter-dominated at a distance $\lesssim 100~R_{\rm S}$.

It should also be noted that we are not taking into account neither the internal energy of the particles nor the conversion of magnetic energy into bulk kinetic energy that accelerates the jet and diminishes even more the magnetization as the jet expands.

\section{Discussion}\label{sec7}

The considerations of the previous section apply only if efficient non-thermal particle acceleration takes place in the hot accreting flow. This is expected to be the case if turbulent magnetic fields are present in the flows, something supported by numerical simulations \cite{Kowal2011,Sironi2014,Lazarianetal2015,Gouveiaetal2019}. In systems where efficient acceleration does not occur in the accretion flow, mass load should exclusively proceed through the only relevant thermal channel: thermal photon annihilation, and also possible enhancements via electromagnetic cascades for very-low luminous systems\footnote{Cascades develop only if the magnetic configuration is such as not to allow catastrophic synchrotron losses, see Section \ref{subsect:cascades}.}. As it is shown in Figure \ref{fig:e_dist}, thermal photo-annihilation is not effective far from the black hole. In the first 100 $R_{\rm S}$ it decreases by more than 5 orders of magnitude. However, at the very base of the jet it can inject about 10 pairs per cubic centimeter and per second. Baryons, on the other hand, will be only injected by direct neutron decay. These neutrons will be the result of inelastic collisions of protons in the innermost part of the accreting flow, with a luminosity not exceeding $10^{40}$ erg s$^{-1}$ (see Figure \ref{fig:neutron_inj}).  They also depend upon the presence of a high-energy proton component within the flow in order to be created with high enough Lorentz factors to reach the funnel. Interestingly, the required Lorentz factor depends on the length scale of the system, namely the Schwarzschild radius. For less massive black holes, neutrons produced by mildly relativistic thermal protons can be able to reach the funnel as well, so neutron effects can be more important in microquasars than in AGNs \cite{Vilaetal2014}. 

The lack of turbulence should affect the entrainment of clumps of matter or the mixing at the boundary layer between the jet and the wind. Consequently, initial jets propagating through purely thermal hot flows should have to acquire a baryon load much farther away, when the jet interacts with different types of obstacles such as stars or clouds \cite{Komissarov1994,Araudoetal2010,Araudoetal2013}.

In the case of hot flows with non-thermal content, the specific initial ratio $R_{e^\pm / p}$ of electrons to protons will depend on the details of the acceleration mechanism in the flow, the energy budget available to the different species of particles, and the details of the losses, but our calculations show that for a range of parameters thought to be typical of these systems the initial particle content will be mostly in the form of electron-positron pairs.

The production rate of pairs in the funnel, as mentioned in Section \ref{sec:photon_ann}, is $\sim$10$^{48}$ s$^{-1}$. This mean a power of $L_{e^{\pm}}\sim 10^{42}$ erg s$^{-1}$. If there is not further mass loading due to entrainment, the bulk Lorentz factor of the jet should be $\Gamma\sim P_{\rm BZ}/L_{e^{\pm}}$. If the spin of the black hole is $a\sim 1$, $B\sim 10^4$ G, and $M_{\rm BH}\sim 10
^9$ $M_{\odot}$, we have $P_{\rm BZ}\sim 10^{46}$ erg s$^{-1}$ from Equation (\ref{PBZ2}). Then, $\Gamma\sim 10^4$. This value is far from the typical value $\Gamma\sim10$ inferred from superluminal motions in AGNs. Since the parameters we have adopted for the non-thermal photon production in the hot disk are already optimistic, we are lead to three possibilities: (1) somehow baryonic matter is injected by instabilities at the base of the jet (see above), (2) most of the magnetic energy is converted into internal energy of the gas (by reconnection and turbulence) or a mixture of both (1) and (2). In such a complex environment as the base of the jet, likely the latter combined case is closer to the real situation.

\section{Conclusions}\label{sec8}

Relativistic jets are created in the magnetized environment of a rotating black hole. The jets initiate as a pure Poynting outflow powered by the black hole ergosphere. At very short distances from the black hole, the jets seem to start radiating as shown by both rapid variability and directed imaging in the case of nearby sources such as M87. Radiation requires the presence of charged particles within the outflow. Since the initial jet is shielded by the strong magnetic fields, it is not clear how the particles are injected in the magnetically dominated region. 

In this article, we have shown that particle interactions in the hot accretion flow that feeds the black hole can produce an outflow of neutral particles that penetrate in the jet collimation funnel and create in situ charged particles, both leptons and hadrons. The main source of electron-positron pairs is photon-annihilation. These photons with energies above 0.5 MeV are radiated from the disk by a~variety of mechanism. The annihilated against softer photons also from the accretion flow. We have presented detailed calculations that discriminate the different contributions, the energy deposited, and the number of particles. Pairs can be additionally injected by neutron decay, $n\gamma$ interactions in the funnel, and the subsequent decays. The origin of these neutrons is in $pp$ and $p\gamma$ collisions within the hot inflow. Neutron decay inside the jet also contributes with protons, which are caught with the outflow and can in turn produce more pairs through Bethe-Heitler and photo-meson interactions. All~this configures a complex picture, with many competing processes. 

All in all, the initial jet is dominated by lepton, which are the more copiously produced particles. Although only about 1\%  of the mass is initially in baryons, the ratio will change by entrainment and jet interactions with different obstacles. The development of the initial jet/hot-wind interactions in MAD and RIAF models remain to be explored.

\vspace{6pt} 



\authorcontributions{Please add.} 

\funding{This research was funded by by the Argentine
agency CONICET (PIP 2014-00338), the agency ANPCyT (PICT 2017-2865) and the Spanish Ministerio de Economía y Competitividad (MINECO/FEDER, UE) under grants AYA2016-76012-C3-1-P and  PID2019-105510GB-C31.}

\acknowledgments{The authors thank Florencia Vieyro for insightful comments, and the three anonymous reviewers for very constructive comments and suggestions that helped to improve the manuscript.}

\conflictsofinterest{The authors declare no conflict of interest.} 


\abbreviations{The following abbreviations are used in this manuscript:\\

\noindent  
\begin{tabular}{@{}ll} 
AGN & Active Galactic Nuclei\\
ADAF & Advection Dominated Accretion Flows\\
BH & Black Hole\\
DSA & Diffusive Shock Acceleration\\
EHT & Event Horizon Telescope\\
FSRQ & Flat Spectrum Radio Quasar\\
\end{tabular}

\noindent  
\begin{tabular}{@{}ll} 
GRMHD & General relativistic magnetohydrodynamics\\
HD& Hydrodynamics\\
ISCO & Innermost Stable Circular Orbit\\
MAD & Magnetically Arrested Disk\\
MHD & Magnetohydrodynamics\\
RIAF & Radiatively Inefficient Accretion Flow\\
SDA & Stochastic Diffusive Acceleration \\
SED & Spectral Energy Distribution \\
SMBH & Super Massive Black Hole 
\end{tabular}}



\reftitle{References}

\end{document}